\documentclass[conference,compsoc]{IEEEtran}
\ifCLASSOPTIONcompsoc
  \usepackage[nocompress]{cite}
\else
  \usepackage{cite}
\fi
\ifCLASSINFOpdf
\else
\fi
\usepackage[utf8]{inputenc}
\usepackage{algorithm}
\usepackage{algpseudocode}
\usepackage{amsfonts,amssymb,amsthm,amsmath}
\usepackage{bbm}
\usepackage{bm}
\usepackage{booktabs} 

\theoremstyle{plain}

\theoremstyle{definition}

\theoremstyle{remark}

\usepackage{array}
\usepackage[caption=false,font=normalsize,labelfont=sf,textfont=sf]{subfig}
\usepackage{textcomp}
\usepackage{stfloats}
\usepackage{url}
\usepackage{verbatim}
\usepackage{graphicx}
\RequirePackage{fix-cm}
\usepackage{listings}

\begin{document}
%
\title{OCELOT: Inference-Leakage Budgets for Privacy-Preserving LLM Agents}

\author{Jin Xie,
        Songze Li
        }
\maketitle

\begin{abstract}
Large language model (LLM) agents increasingly act on a user's behalf-- reading personal files, calling tools, transacting with external services---possibly leaking personally identifiable information (PII) across trust boundaries at every step. Privacy here is a property not of a single output but of an entire trajectory, and three properties make it hard: leakage is cumulative, as individually innocuous releases accumulate across honest-but-curious or colluding sinks into inferences about a protected secret; bidirectional, as a malicious observation can inject instructions that turn the agent's own reasoning model against the user; and task-dependent, as the same field is necessary for one recipient yet gratuitous for another. Per-release contextual-integrity filters, information-flow controls, and posterior-leakage monitors each address part of this but none controls cumulative, inference-based leakage at runtime. We recast agent privacy as \emph{posterior-risk control} and present OCELOT, a runtime mediator that budgets how much an adversary's belief about a secret may improve across a trajectory, rather than filtering outputs. Its mechanism, \emph{Witness-Verified Declassification}, separates judgment from trust: an untrusted, locally fine-tuned defender model inspects each candidate release and emits structured evidence---labeled atoms and proposed declassification operators---which a deterministic verifier audits, charging a certified min-entropy cost for the chosen variant and authorizing the least-disclosing useful release under a sink-trust-weighted budget recorded on a tamper-evident ledger. Because every unverified atom is costed at worst-case risk, a jailbroken defender yields conservative fallback, never silent leakage. This gives a two-layer guarantee: budget non-exceedance holds deductively, independent of any LLM, while semantic privacy holds empirically, to the degree calibrated costs upper-bound true adversarial inference. Across diverse agent benchmarks and recent defenses, OCELOT attains significantly lower leakage at higher task utility, resists adaptive injection, jailbreak, cumulative inference, and sink collusion, and adds only modest overhead.
\end{abstract}

\IEEEpeerreviewmaketitle

\section{Introduction}
\label{sec:intro}

An LLM agent that books a flight does not merely answer a
question; it \emph{acts}. To finish the task it reads the user's
saved profile, queries a travel broker for fares, submits a
passenger record to an airline, and authorizes a charge with a
payment processor---disclosing a different slice of the user's
identity to a differently-trusted party at each step. This
action-taking posture, standard in production agent harnesses
(Claude Code, OpenAI Codex, Cursor)~\cite{yao2022react,young2025effective, lopopolo2026harness,pan2026naturallanguageagentharnesses},
is what makes agents useful, and a new privacy liability: an agent
is a conduit through which personally identifiable information
(PII) flows continuously, across trust boundaries, on the user's
behalf.

Prior work on language-model privacy centers on what a
\emph{model} memorizes or what a \emph{single} output
reveals~\cite{carlini2021extractingtrainingdatalarge,staab2024memorizationviolatingprivacyinference}. Agent
privacy is a different object: exposure is a property of an entire
trajectory of releases to many sinks, and what matters is not
whether any message contains an identifier but how much an
adversary observing the trajectory can \emph{infer} about a
protected secret. A defense that inspects releases in isolation
cannot see this---recent multi-agent benchmarks confirm the
dominant residual leakage is accumulated and indirect, not any
single offending message~\cite{yagoubi2026agentleakfullstackbenchmarkprivacy}.

\begin{figure}[t]
\centering
\includegraphics[width=0.98\columnwidth]{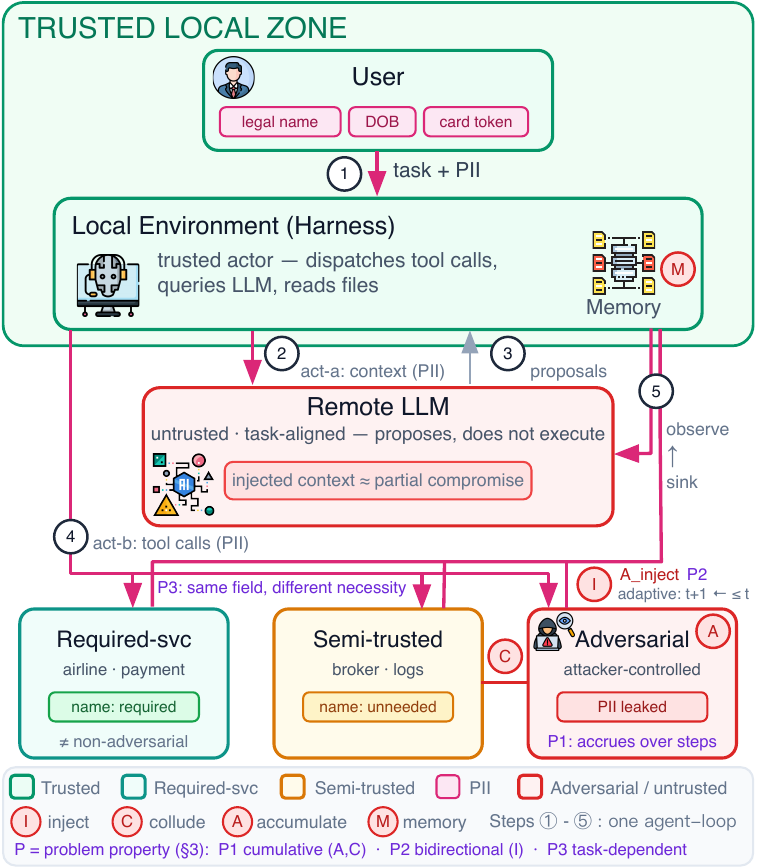}
\vspace{-3mm}
\caption{Privacy exposure in one LLM-agent loop: a user's PII
(pink) crosses trust boundaries to the remote LLM (act-a) and to
external sinks (act-b), with observations returning inward
(observe). Three properties (\S\ref{sec:problem}) defeat
per-release defenses---cumulative accrual and collusion (P1,
glyphs \textsf{A}/\textsf{C}), bidirectional injection (P2,
\textsf{I}), and task-dependent necessity (P3); unverifiable sinks
default to \textsc{adversarial}.}
\label{fig:threat}
\vspace{-5mm}
\end{figure}

Three properties, traced in Figure~\ref{fig:threat}, make agent
privacy mediation hard. \textbf{(P1) Leakage is cumulative:}
individually innocuous releases combine---at one
honest-but-curious sink, or across colluding sinks---to
reconstruct a secret no release named. \textbf{(P2) Leakage is
bidirectional:} information escapes when the agent ingests
observations, not only when it acts; a tool can return content
that injects instructions, subverting the task-aligned but
untrusted reasoning model into leaking more on later
turns~\cite{debenedetti2024agentdojodynamicenvironmentevaluate}. \textbf{(P3) Leakage is
task-dependent:} the same field can be required by one sink and
gratuitous for another in the same trajectory, so no universal
safe answer exists. Together these defeat the per-release, binary,
and monitoring defenses deployed today.

The question a mediator must answer at runtime is therefore not
``allow or deny?'' but: \emph{given the task, a candidate release,
the destination sink, the trajectory so far, and the remaining
privacy budget, what is the least-disclosing useful variant that
can be authorized with auditable evidence?} This demands three
things no prior defense provides together: a quantitative,
composable notion of leakage that accumulates across a trajectory;
a mechanism that chooses among semantic-granularity variants
rather than a binary verdict; and an authorization decision whose
soundness does not depend on the correctness of any LLM. Existing
runtime defenses each supply one of these and miss the rest
(\S\ref{sec:related}).

We present \textbf{OCELOT} \footnote{\textbf{OCELOT}:
\emph{Operator-Calibrated, Evidence-verified, Leakage-budgeted Online
Trajectory mediator.}} , a runtime mediator that recasts agent
privacy as \emph{posterior-risk control}: it budgets how much an
adversary's posterior over a secret may improve across a
trajectory, rather than filtering outputs. Its mechanism,
\emph{Witness-Verified Declassification} (WVD), resolves the
central tension---semantic judgment needs an LLM, yet no LLM can be
trusted in an adversarial loop---by separating the two. An
untrusted, locally fine-tuned \emph{defender} applies a versioned
rubric to each candidate release and emits structured
\emph{evidence}: labeled atoms and a proposed sequence of
declassification operators. A deterministic, trusted
\emph{verifier} audits this evidence, computes a certified
min-entropy cost for the proposed variant from calibrated
per-operator bounds, and authorizes the least-disclosing useful
variant under a sink-trust-weighted budget recorded in a
Merkle-chained ledger. The defender is
\emph{honest-but-jailbreakable}: because the verifier costs any
unlabeled or mislabeled atom at top risk, a coerced defender
yields \emph{fallback}, never silent leakage. The result is a
two-layer guarantee we state honestly---budget non-exceedance
holds \emph{deductively} by ledger arithmetic, independent of any
LLM, while the semantic-privacy implication holds
\emph{empirically}, to the extent calibrated costs upper-bound true
adversarial inference, which we measure rather than assume.

\smallskip
\noindent\textbf{Contributions.}
\begin{itemize}
\item \emph{Problem and goal.} We formalize privacy mediation for
LLM agents around properties P1--P3 and a two-layer security goal
separating a deductive budget guarantee from an empirical,
calibration-relative privacy claim (\S\ref{sec:problem}).
\item \emph{Mechanism.} We design WVD: an untrusted defender
producing verifiable evidence under grammar-constrained decoding,
and a deterministic verifier that authorizes least-disclosure
variants via a four-phase procedure with conservative defaults,
structural-field obligations, and a Merkle-chained leakage ledger
(\S\ref{sec:method}).
\item \emph{Costs and training.} We calibrate per-operator
min-entropy costs as adversarial-inference upper-confidence bounds,
audit them with a bounded-treewidth credal circuit, and train the
defender with a GRPO reward computed by the verifier itself, so the
policy optimizes against the exact deployment-time check
(\S\ref{sec:method:algebra},~\S\ref{sec:method:defender}).
\item \emph{Benchmark and protocols.} We release OCELOT-Traj-500
and three pre-registered protocols---calibration, escape-rate, and
rubric construction---for reproducible evaluation
(Appendices~\ref{app:prereg-a}--\ref{app:dataset}).
\item \emph{Evaluation.} Across five benchmarks and five recent
defenses, OCELOT attains the lowest leakage \emph{and} highest task
success on every benchmark, records zero budget exceedances over
$22{,}585$ trajectory-instances, and degrades far more gracefully
than every non-denial baseline under injection, jailbreak,
cumulative inference, and sink collusion---at $18\%$ wall-clock
overhead (\S\ref{sec:exp}).
\end{itemize}

\section{Related Work}
\label{sec:related}

Production LLM agents run a harness that brokers all I/O under
permissions~\cite{lopopolo2026harness,young2025effective}, and their dominant
risk is a task-aligned model executing instructions injected into
external content; AgentDojo~\cite{debenedetti2024agentdojodynamicenvironmentevaluate} systematizes such
attacks and serves as our injection benchmark, and the structural
defense it motivates---stripping authority from untrusted data
rather than detecting it---is one we reuse on the observe channel.
The privacy threat is subtler. Beyond verbatim
memorization~\cite{carlini2021extractingtrainingdatalarge}, LLMs can \emph{infer}
sensitive attributes from benign text~\cite{staab2024memorizationviolatingprivacyinference}, so
the quantity that matters is not whether a message contains an
identifier but how much an adversary can infer across a
trajectory. ConfAIde~\cite{mireshghallah2024llmssecrettestingprivacy} casts contextual
integrity~\cite{nissenbaum2004privacy} as tiered single-release
judgments, whereas AgentLeak-2026~\cite{yagoubi2026agentleakfullstackbenchmarkprivacy} shows that
in multi-agent trajectories the dominant residual leakage is
cumulative and indirect; we evaluate on both, using single-release
ConfAIde to stress-test rather than flatter our trajectory
framing.

Runtime defenses fall into three families, and our baselines span
all three. Per-release contextual-integrity
filters---FlexGuard~\cite{ding2026flexguardcontinuousriskscoring} and
PrivacyChecker~\cite{wang2025privacyactionrealisticprivacy}---judge whether each
disclosure is appropriate but keep no cross-trajectory state, so
independently ``appropriate'' releases can jointly reconstruct a
secret (P1), and their binary verdicts cannot trade granularity
for utility (P3). Information-flow and capability
systems---CaMeL~\cite{debenedetti2025defeatingpromptinjectionsdesign} and RTBAS~\cite{zhong2025rtbasdefendingllmagents}---gate flows
across trust boundaries with strong worst-case safety but answer
only \emph{whether} information may flow, not \emph{how much}: a
field is released at full fidelity or denied, the characteristic
IFC trade of low leakage bought with heavy over-denial, which
makes CaMeL our denial-based reference on the privacy axis.
Closest to our cost model, AGENTDAM~\cite{zharmagambetov2026agentdam} estimates how
much a release improves an adversary's posterior but is a
\emph{monitor}---it quantifies risk without enforcing a budget and
forwards injected observations unmediated; we compare against both
its monitoring configuration and a steelmanned
threshold-and-block variant (Appendix~\ref{app:baselines}), yet
thresholding a per-release estimate still cannot bound a cumulative
sum. Training-time regularization ~\cite{asif2026informationtheoreticprivacycontrolsequential} bounds
leakage during fine-tuning and is complementary but orthogonal,
requiring weight access and offering no runtime control over which
field reaches which sink at what granularity.

Our mechanism builds on established theory: per-operator costs are
min-entropy leakage and Bayes-vulnerability bounds from
quantitative information flow~\cite{smith2009foundations,kopf2007information},
our trajectory accounting parallels a privacy odometer in
differential-privacy composition~\cite{dwork2014algorithmic}, the
independent audit is a credal sum-product
circuit~\cite{darwiche2009modeling,maua2014probabilistic}, and rubric
obligations are discharged by an SMT
solver~\cite{barrett2010smt,de2008z3}. The contribution is
their synthesis into a runtime mediator whose budget guarantee is
independent of LLM behavior. Each prior family supplies one
ingredient; to our knowledge OCELOT is the first runtime agent
privacy defense to provide all three at once---\emph{controlling}
rather than merely measuring cumulative inference risk, acting at
\emph{semantic granularity} rather than allow/deny, and grounding
authorization in a \emph{deterministic verifier} whose guarantee
does not depend on any LLM.

\section{Problem Formulation}
\label{sec:problem}

\subsection{The Privacy-Mediation Problem}
\label{sec:problem:setting}

An LLM agent executes a user-issued task as a multi-step trajectory:
it reads observations from tools and the environment, maintains memory
across turns, and emits actions to external sinks --- APIs, logs,
third-party services, and other agents. Each step may cross a trust
boundary, so a privacy defense must answer, at runtime, what may be
released to whom, in what form, and at what cost. We call this the
\emph{privacy-mediation problem for LLM agents} and identify three
properties distinguishing it from problems prior defenses address.

\smallskip
\noindent\textbf{(P1) Leakage is cumulative.}
Individual releases may carry no explicit identifier, yet several
weak releases let a sink (or colluding set of sinks) infer a
protected secret. Per-release filters cannot detect this, and recent
benchmarks confirm that the dominant residual leakage in multi-agent
systems flows through accumulated, indirect
disclosures~\cite{yagoubi2026agentleakfullstackbenchmarkprivacy}.

\smallskip
\noindent\textbf{(P2) Leakage is bidirectional.}
Information escapes not only when the agent acts toward a sink but
also when it ingests observations: a tool may return content crafted
to extract disclosure on later turns, or to inject instructions that
cause subsequent releases to leak
more~\cite{debenedetti2024agentdojodynamicenvironmentevaluate}.

\smallskip
\noindent\textbf{(P3) Leakage is task-dependent.}
The same field can be \emph{necessary} for one sink and
\emph{unnecessary} for another in the same trajectory: a passenger's
date of birth is needed by an airline at ticket issuance but not by a
flight-search broker. No universal safe answer exists; the question
is what disclosures are necessary for \emph{this} task, at \emph{what}
granularity, to \emph{which} sink.

\smallskip
Prior defenses miss (P1)--(P3). IFC- and capability-style agent
defenses~\cite{debenedetti2025defeatingpromptinjectionsdesign,zhong2025rtbasdefendingllmagents}
decide whether information may flow, not how much or at what
granularity. Contextual-integrity
checkers~\cite{wang2025privacyactionrealisticprivacy,ding2026flexguardcontinuousriskscoring}
judge appropriateness one release at a time, without cumulative
accounting. Posterior-leakage
estimators~\cite{zharmagambetov2026agentdam} measure leakage but do
not control it. Information-theoretic training-time
regularization~\cite{asif2026informationtheoreticprivacycontrolsequential}
bounds leakage via fine-tuning, requiring weight access and offering
no runtime mediation. None jointly answer:

\begin{quote}
\emph{Given the current task, candidate release, sink, trajectory
history, and remaining privacy budget, what is the least-disclosing
useful variant of the release that can be authorized with auditable
evidence?}
\end{quote}

\noindent
Answering it requires a quantitative composable notion of leakage that
accumulates across a trajectory, a choice among semantic-granularity
variants rather than binary allow/deny, and an authorization whose
soundness does not depend on any LLM.

\subsection{System Setting and Information Flow}
\label{sec:problem:system}

We consider an LLM-agent system following the harness architecture of
production agents (Claude Code, OpenAI Codex, Cursor,
OpenHarness)~\cite{young2025effective,lopopolo2026harness}. Four
parties are relevant; the mediator this paper introduces is a fifth
component inside one of them.

\smallskip
\noindent\textbf{User.} Issues a task and trusts the local
environment. Holds a typed secret $S$ with a per-task prior $P(S)$,
default uniform.

\smallskip
\noindent\textbf{Local environment (harness).} Trusted software on
the user's machine. It brokers all communication --- reading local
files, managing persistent memory, querying the remote LLM,
dispatching tool calls, handling permissions and sandboxing --- and
is the \emph{actor} in the agent loop: the LLM only proposes; the
harness executes.

\smallskip
\noindent\textbf{Remote LLM.} The reasoning component, reached over an
API (e.g., \texttt{/v1/messages},
\texttt{/v1/responses})~\cite{lopopolo2026harness}. It receives task
context, prior observations, and tool schemas and returns text,
tool-call proposals, and reasoning; it does not execute. It is
task-aligned but \emph{untrusted}: content injected into its context,
typically through observations returned by sinks, can subvert its
behavior~\cite{lopopolo2026harness,debenedetti2024agentdojodynamicenvironmentevaluate}.

\smallskip
\noindent\textbf{External sinks.} Tools, APIs, websites, MCP servers,
logs, memory stores, and other agents the harness contacts on the
user's behalf. Each carries a declared trust class:
\textsc{adversarial} (under attacker control); \textsc{semi-trusted}
(honest-but-curious, may log but not actively attack); and
\textsc{required-service} (an authenticated endpoint, e.g., a payment
processor, whose disclosure is required for the task). The classes are
policy weights, not safety assumptions: \textsc{required-service} does
\emph{not} mean non-adversarial, and unverifiable sinks default to
\textsc{adversarial}.

\smallskip
\noindent\textbf{Local mediator.} The component this paper introduces,
inside the harness: a local fine-tuned LLM (the defender,
\S\ref{sec:method:defender}) and a deterministic verifier
(\S\ref{sec:method:verifier}), intercepting every information flow
crossing the harness boundary toward the remote LLM or external sinks.

\smallskip
\noindent\textbf{Trust-asymmetric channels.}
Information crossing the harness boundary divides into four channels,
distinguished by the trust relation between source and destination
rather than by topology:

\begin{itemize}
  \item \textbf{Act channel.} Trusted local content flowing outward, in
        two legs: \emph{harness $\to$ remote LLM} (task context, file
        contents, prior observations sent to the API) and \emph{harness
        $\to$ external sink} (tool calls).
  \item \textbf{Observe channel.} Content from external sinks flowing
        inward through the harness to the remote LLM.
  \item \textbf{Memory channel.} Trajectory state written to a
        persistent store, readable by future trajectories with possibly
        different trust.
  \item \textbf{User channel.} Confirmations and disclosures back to the
        user; audited but unbudgeted.
\end{itemize}

The four channels aggregate to the seven topological channels
of~\cite{yagoubi2026agentleakfullstackbenchmarkprivacy}, which we use
only for benchmark comparison.

\subsection{Adversary Model}
\label{sec:problem:adversary}

\noindent\textbf{Capability.}
The adversary controls every \textsc{adversarial}-class sink and may
inject content into any observation reaching the remote LLM through
the harness. It observes every act-channel release to
\textsc{adversarial} and \textsc{semi-trusted} sinks and every
memory-channel write later readable by such sinks, and may collude
across them. It is \emph{adaptive}: injected content on turn $t{+}1$
may depend on releases observed up to turn $t$. No compute bound is
assumed; inference may combine released variants with arbitrary public
background knowledge. We do not assume the provider's infrastructure is
compromised, but a remote LLM whose context contains injected
instructions behaves indistinguishably from a partly-compromised one,
and the defense must account for this.

\smallskip
\noindent\textbf{Goal.}
The adversary aims to reduce the min-entropy of $S$ given its view
below a target threshold; equivalently, to drive
$\Pr[S{=}s \mid \mathrm{view}]$ above a threshold for some $s$.

\smallskip
\noindent\textbf{Component trust.}
The harness and the mediator inside it are trusted at the level of
standard system integrity (no kernel compromise, no key exfiltration);
orthogonal hardening is out of scope. The remote LLM is \emph{untrusted
but task-aligned}: its proposed tool calls are inputs to mediation, not
authorizations. The defender is also an LLM and is modeled as
\emph{honest-but-jailbreakable}: its weights are trusted
(operator-controlled fine-tuning), but its outputs are not, since
injection in its input may with bounded probability coerce misleading
content. The mediator therefore treats the defender as a constrained
witness producer under deterministic verification, not a classifier
whose verdict is dispositive.

\smallskip
\noindent\textbf{Out of scope.}
A compromised user; compromise of the harness's runtime integrity
(orthogonal TEE/signed-binary infrastructure); compromise of the LLM
provider's infrastructure or model weights; physical-layer side
channels; denial-of-service against the mediation pipeline; and
inference attacks exploiting \emph{only} public background knowledge
without combining it with any released variant. The last exclusion is
consistent with the unbounded priors granted above: a charged leakage
is one in which some released variant moves the posterior.

\subsection{Security Goal}
\label{sec:problem:goal}

We formulate the security goal in two regimes --- deductive (Layer~1a) and empirical (Layers~1b and~2) --- matching what a defense can honestly guarantee.

\smallskip
\noindent\textbf{Layer 1a (deductive): budget non-exceedance.}
For every trajectory of length $k$ it authorizes and every adversary
in the class above, the system must enforce
\begin{equation}
  \sum_{i=1}^{k} \mathrm{cost}_i \;\le\; B,
  \label{eq:security-goal}
\end{equation}
where $\mathrm{cost}_i$ is the certified release cost charged at step
$i$ (defined in \S\ref{sec:method:verifier}) and $B$ is a fixed per-task base budget. Sink trust enters through $\mathrm{cost}_i$ via the trust multiplier $\mu(z)$ (\S\ref{sec:method:verifier}), not through $B$. This holds without exception, by ledger arithmetic (Theorem~SP-1).

\smallskip
\noindent\textbf{Layer 1b (model-relative): cost soundness.}
For every authorized release with variant $v$, the
\emph{model-relative} leakage $\widehat{L}(v;S)$ --- the min-entropy
reduction computed under the calibration model of Pre-Reg.~A ---
satisfies $\widehat{L}(v;S)\le\mathrm{cost}$, except with empirically
bounded probability $\delta$: the residual probability that the
mediator's LLM-based components emit labels passing the verifier's
validate phase while misclassifying an atom's field outside the
verifier-considered field set (measured in Pre-Reg.~B,
\S\ref{app:prereg:b}).

\smallskip
\noindent\textbf{Layer 2 (model-relative): semantic privacy.}
\emph{If, on the released variants, $\widehat{L}$ upper-bounds the true
adversary's realized MAP advantage, then for every authorized trajectory}
\begin{equation}
  H_\infty(S) - H_\infty(S \mid \mathrm{view}) \;\le\; B,
  \label{eq:layer2}
\end{equation}
\emph{where $\mathrm{view}$ is the adversary's realized observation and
$H_\infty(S\mid\mathrm{view}) = -\log_2 \max_s \Pr[S{=}s\mid\mathrm{view}]$
is the realized (not channel-averaged) conditional min-entropy; the
left-hand side is then the realized MAP advantage accumulated over the
trajectory --- the quantity the per-release costs bound.}

\smallskip
\noindent
The split is deliberate: \eqref{eq:security-goal} holds by ledger
arithmetic regardless of how $\widehat{L}$ relates to the true
posterior, whereas the Layer-2 implication is an empirical claim, since
no defense can deductively bound leakage against an unbounded-compute
adversary with arbitrary background knowledge. A credible defense must
therefore confine the LLM-dependent assumptions to the calibrated
terms. The remainder of this paper presents such a defense.

\section{Witness-Verified Declassification}
\label{sec:method}

\begin{figure*}[t]
\centering
\includegraphics[width=0.95\textwidth]{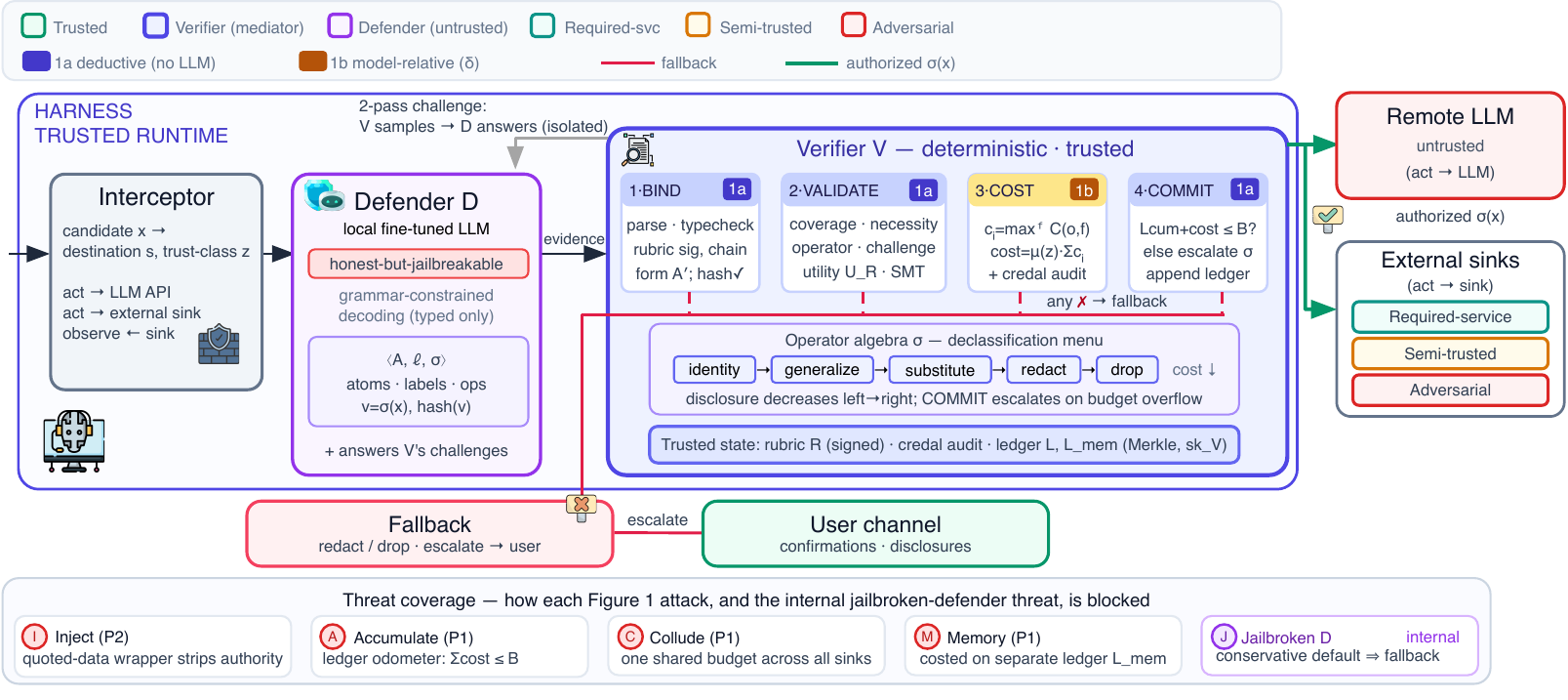}
\vspace{-4mm}
\caption{OCELOT architecture. For each candidate release, the
untrusted defender $D$ emits evidence $\langle A,\ell,\sigma\rangle$;
the deterministic verifier $V$ audits it in four phases against its
trusted state and authorizes the least-disclosing variant $\sigma(x)$
or falls back. Bind, Validate, and Commit are deductive (no model
dependence); Cost is model-relative (residual $\delta$). The bottom
band maps each Figure~\ref{fig:threat} threat
(\textsf{I},\textsf{A},\textsf{C},\textsf{M}) and the internal
jailbroken-defender threat (\textsf{J}) to its defense; any failed
check yields fallback, never silent release.}
\label{fig:architecture}
\vspace{-5mm}
\end{figure*}

This section presents \textbf{OCELOT}, a runtime privacy mediator realizing the security goal of~\S\ref{sec:problem:goal} under the adversary model
of~\S\ref{sec:problem:adversary}. Its mechanism, \emph{Witness-Verified
Declassification} (WVD), pairs an untrusted local LLM that applies a
versioned rubric to each candidate release --- emitting structured atom
labels and an operator proposal --- with a deterministic verifier that
checks the labels against rubric obligations, computes a certified
leakage cost, and authorizes the least-disclosing useful variant under
a per-trajectory, sink-trust-weighted budget. Figure~\ref{fig:architecture}
shows the architecture; Algorithm~\ref{alg:wvd} the authorization
procedure for one mediated release.

\subsection{Architecture and Rubrics}
\label{sec:method:overview}

The defender $D$ (\S\ref{sec:method:defender}) and verifier $V$ run
inside the harness; $V$ holds a versioned rubric library $\mathcal{R}$,
a per-trajectory Merkle-chained ledger $\mathcal{L}$, a credal
posterior audit engine, and a signing key $sk_V$.

\smallskip
\noindent\textbf{What OCELOT mediates.}
The three boundary-crossing flows of~\S\ref{sec:problem:system} (act
legs a and b, and the observe channel) are mediated by one procedure:
whenever the harness is about to send a message across the boundary in
either direction, $V$ intercepts it before transmission. We write $x$
for the candidate release, $d$ for the destination (an external sink or
the remote LLM endpoint), $z$ for its declared trust class, and $S$ for
the user's secret. Algorithm~\ref{alg:wvd} summarizes the procedure;
its symbols are defined as they arise below.

\begin{figure}[htbp]
\vspace{-3mm}
\begin{algorithm}[H]
\small
\caption{WVD authorization for one candidate release.
Lines~1--3 resolve the rubric set and obtain the defender's evidence
(two isolated passes); lines~4--6 are \textsc{Bind}, line~7
\textsc{Validate}, lines~8--9 \textsc{Cost} (including the credal
audit), and lines~10--11 \textsc{Commit}. Named procedures are defined
in surrounding prose.}
\label{alg:wvd}
\begin{algorithmic}[1]
\Require candidate $x$, destination $d$ with trust class $z$,
         ledger $\mathcal{L}$
\State $R \gets \textsc{ResolveRubricSet}(x, d)$
       \Comment{set-valued; conservative join}
\State $(A, \ell, \sigma) \gets D.\textsc{ApplyRubric}(x, R)$
\State $c^{\text{ch}} \gets D.\textsc{AnsChallenges}(
       V.\textsc{SampleChallenges}(R), x)$
       \Comment{second isolated pass; clauses chosen by $V$}
\State $A' \gets A \cup \textsc{StructuralFieldAtoms}(x, R)$
\State $\ell' \gets \textsc{ApplyConservativeDefault}(\ell, A')$
\If{$\neg\textsc{Bind}(x, \sigma, R, \mathcal{L})$}
  \Return \textsc{fallback}
\EndIf
\If{$\neg\textsc{Validate}(\ell', c^{\text{ch}}, \sigma, A', R, U_R)$}
  \Return \textsc{fallback}
\EndIf
\State $\mathrm{cost} \gets \textsc{Cost}(\sigma, A', \ell', R,
       \mu(z))$
\If{$\textsc{Audit}(\sigma(x)) > \mathrm{cost} + \varepsilon(R)$}
  \Return \textsc{fallback}
\EndIf
\If{$L_{\text{cum}} + \mathrm{cost} > B$}
  \Return \textsc{EscalateAndRetry}($\sigma$, $R$)
\EndIf
\State $\mathcal{L}.\textsc{Append}(e_i)$;\quad
       \Return $\sigma(x)$
\end{algorithmic}
\end{algorithm}
\vspace{-9mm}
\end{figure}

\smallskip
\noindent\textbf{Division of labor.}
Open-vocabulary semantic classification --- ``does this span imply a
sensitive attribute?'' --- is a task LLMs do well and deterministic
code does not, so the defender atomizes and labels while the verifier
audits: checking the atom set against rubric coverage obligations and
structural-field obligations from trusted tool schemas, verifying
label-operator consistency, and computing certified costs. Omission
attacks --- a malicious defender silently dropping a sensitive atom ---
are countered by four mechanisms in concert: rubric coverage
obligations (below), structural-field obligations from trusted tool
schemas, adversarial omission-curriculum training
(\S\ref{sec:method:defender}), and two-pass consistency challenges
(\S\ref{sec:method:defender}). None is individually airtight; their
combination makes the architecture defensible.

\smallskip
\noindent\textbf{Rubric resolution is set-valued.}
A structured task typer in $V$'s trusted code extracts a tuple
$(\text{action\_category},\,\{\text{field\_types}\},\,
\{\text{sink\_trust\_class}\})$ from \emph{trusted structured metadata
only}: tool schemas, endpoint identity, signed registry labels. The
LLM API endpoint is itself a destination with its own declared trust
class; free-form user text contributes candidate tags but cannot
select a less restrictive rubric on its own. The tuple keys a lookup
returning the set $R$ of rubrics matching the structural signature.
Under ambiguity (multiple matches), $V$ enforces the intersection of
admissible operators, the maximum of per-(operator, field) costs, and
the minimum of base budgets. Under no match, a fallback rubric
$r_{\bot}$ applies, admitting only \textsc{redact} and \textsc{drop}
with a small budget. Adversarial task phrasing therefore cannot
increase disclosure.

\smallskip
\noindent\textbf{What a rubric contains.}
A rubric $r \in \mathcal{R}$ for one task category is a typed, signed
record with seven components: (1) the \emph{task signature} (action
category, field types, destination trust class) keying lookup; (2) the
\emph{secret-field schema} drawing from a global field ontology (e.g.,
\textsc{legal-name}, \textsc{dob}, \textsc{payment-token}); (3) \emph{coverage obligations} --- per-field
trigger predicates (regex over structured schemas, named-entity-type
constraints, lexicon matches) the verifier evaluates over the
candidate, declaring that any candidate span matching a predicate must
correspond to an atom with a non-\textsc{none} field label; (4)
\emph{necessity rules} --- per-(field, destination-class) clauses
declaring admissible necessity levels; (5) the \emph{operator algebra}
of Table~\ref{tab:operators} with per-(operator, field) calibrated
costs $C(\text{op}, f)$, the generalization level $k$ for
\textsc{generalize}, and the template-ID list for \textsc{substitute};
(6) \emph{budget and trust weights} --- per-trajectory base budget
$B_0$, trust-class multiplier $\mu(z)$ ($\mu(z)\ge 1$, so the budget
never under-charges leakage), and calibration margin $\varepsilon(R)$;
(7) the \emph{utility predicate} $U_R$ over the proposed variant,
evaluated deterministically by $V$ on required-field atoms. Coverage
obligations do not constitute semantic completeness; they specify the
minimal flagging discipline $V$ enforces.

\smallskip
\noindent\textbf{How rubrics are built.}
Rubric authoring is a pre-registered three-stage offline process
(Pre-Reg.~C): domain-expert \emph{drafting} against a template
(optionally using a frontier model as a assistant rather than
an authority); \emph{validation} against a held-out corpus of $\geq
100$ realistic agent trajectories per task category, scoring
coverage-obligation hit rate, necessity-rule annotator agreement, and
operator-algebra completeness; and \emph{adversarial probing}, a
red-team pass that attempts to construct releases satisfying all
coverage obligations yet leaking, with failures triggering rubric
patches and re-validation. The result is not semantic completeness but
a transparent, versioned, externally auditable artifact.

\smallskip
\noindent\textbf{Running example.}
A user instructs the harness: \emph{``Find a flight from San Francisco
to Boston tomorrow morning under \$400 and buy it with my saved
card.''} Legal name, DOB, and payment token are typed secret fields;
the legal name is unnecessary for a flight-search broker yet required
by the airline at ticket issuance, and the payment token must reach
only the processor. We trace the full trajectory at the end
of \S\ref{sec:method:verifier}.

\subsection{The Defender: Role and Training}
\label{sec:method:defender}

The defender is a fine-tuned LLM whose only role is to apply the
matched rubric set $R$ to a candidate $x$, producing atoms, labels,
and an operator proposal. $D$ emits only well-typed objects under
grammar-constrained
decoding~\cite{melcer2024constraineddecodingfillinthemiddlecode};
free-form text, authorization verdicts, and out-of-schema content are
structurally unrepresentable. The defender's capability bounds the
system's \emph{utility} (failing to identify a span causes fallback)
but not its \emph{safety} (mislabeling causes fallback, not unchecked
release); the training pipeline below optimizes the former, the
verifier of~\S\ref{sec:method:verifier} enforces the latter.

\smallskip
\noindent\textbf{The atom-label-proposal object.}
For each candidate, $D$ produces a triple $(A, \ell, \sigma)$, where
$A$ is a set of contiguous candidate spans tagged with structural
role; $\ell$ is the per-atom label vector over $A$; and $\sigma =
(o_1, \ldots, o_n)$ is a sequence of operators from the algebra
of~\S\ref{sec:method:algebra}, applied to designated atoms, yielding
the proposed variant $v = \sigma(x)$ accompanied by $\text{hash}(v)$.
Each label in $\ell$ asserts four attributes of its atom $a$: (i) the
secret field $f$ touched by $a$ (\textsc{none} if non-sensitive); (ii)
the relation between $a$ and $f$, drawn from $\{\textsc{direct},
\textsc{implicit}, \textsc{inferential}\}$; (iii) the necessity level
of $a$ for the destination trust class $z$, drawn from
$\{\textsc{required}, \textsc{helpful}, \textsc{unnecessary}\}$; and
(iv) the supporting rubric clause identifier.

The verifier augments $A$ with atoms required by the rubric's
structural-field obligations to form $A' = A \cup
\textsc{StructuralFieldAtoms}(x, R)$, and assigns the
\emph{conservative default} \textsc{unknown-sensitive/top-risk} to any
atom in $A' \setminus A$ or to any atom in $A$ with a malformed label.
Atoms under the conservative default admit only \textsc{drop} or
\textsc{redact} unless a trusted rubric clause establishes necessity.
On the observe channel, $D$ additionally labels each inbound atom
\textsc{task-relevant} or \textsc{private-content}, and every label and
operator carries a provenance tag (trajectory step, tool observation
[untrusted], user input [trusted], or rubric clause).

\smallskip
\noindent\textbf{Two-pass consistency challenges.}
After receiving $\ell$, $V$ --- not $D$ --- samples randomized
challenge clauses from a per-rubric pool (sealed at rubric
construction, Pre-Reg.~C): predicates of the form ``does atom $a$ touch
field $f$?'' or ``is the necessity of $a$ consistent with rubric clause
$c$?''. $D$ answers them in a second, isolated decoding pass with no
carry-over context, and $V$ checks inter-pass consistency. Because $V$
chooses the clauses, a compromised $D$ cannot pre-select a
self-consistent challenge set. Challenges are an \emph{empirical
evasion detector}, not a soundness mechanism --- a self-consistent
compromised defender can pass both passes --- so their role is to
estimate and reduce the residual escape rate $\delta$
(\S\ref{sec:problem:goal}; quantified in Pre-Reg.~B), not to establish
safety, which rests on conservative defaults and cost upper bounds.

\smallskip
\noindent\textbf{Synthetic data generation for SFT.}
We construct supervised-fine-tuning data using the rubrics as ground
truth, avoiding the impractical scale of human-labeled trajectories.
The pipeline has three stages. \emph{(i) Trajectory sampling}
instantiates each rubric's task signature across secret-field values
from its declared populations and generates tool-call payloads with a
frontier model, yielding candidates with known field-to-span truth.
\emph{(ii) Rubric-application labeling} has a frontier model (a
labeling oracle, not a runtime authority) produce the $(A, \ell,
\sigma)$ triple, filtered against the sampling-derived truth with
misalignments corrected or discarded. \emph{(iii) Adversarial
augmentation} adds three pipelines: \emph{omission} embeds or
distributes sensitive spans in obscure phrasing, with targets forcing
the defender to flag them; \emph{injection} synthesizes observe-channel
observations carrying injection-styled instructions, with targets
forcing the \textsc{task-irrelevant} private-content label rather than
compliance; and \emph{distribution shift} produces candidates that
lexically resemble training examples but invert the necessity label
(e.g., DOB in flight search vs.\ ticket purchase), forcing attention to
task context over surface form. The corpus size is declared in
Pre-Reg.~A; SFT proceeds under standard instruction-tuning losses.

\smallskip
\noindent\textbf{Stage~II: GRPO with verifier-conformance reward.}
SFT produces a defender that applies rubrics syntactically but not
always optimally under the authorization procedure. We then apply
Group Relative Policy Optimization
(GRPO)~\cite{shao2024deepseekmathpushinglimitsmathematical} with a
reward computed by the verifier itself. For a defender output $o$ on
candidate $x$,
\begin{equation}
  \begin{aligned}
  R(x, o) \;= & \; \bm{1}\!\left[\text{authorized}(o)\right] \cdot
                \max\!\big(B - \text{cost}(o),\, 0\big)\\
              & \;-\; \lambda \cdot
                \bm{1}\!\left[\text{fallback}(o)\right],
\end{aligned}
  \label{eq:grpo-reward}
\end{equation}
where $\bm{1}[P]$ is the indicator of $P$; $B$ is the per-trajectory
budget; $\text{cost}(o)$ is the WVD cost on $o$; $\text{authorized}(o)$
and $\text{fallback}(o)$ are determined by running the full WVD
procedure on $o$ against a frozen rubric set; and $\lambda > 0$ is the
fallback penalty. The reward is \emph{exact} rather than a learned
approximation --- the verifier is the ground truth for authorization,
so the policy optimizes directly against the deployment-time check ---
and it \emph{generalizes across rubrics}: because the verifier is
rubric-agnostic, one GRPO run trains the defender library-wide without
per-rubric retraining.

\subsection{The Operator Algebra and Calibration}
\label{sec:method:algebra}

\begin{table}[t]
\small
\centering
\caption{The five WVD operators. We omit \textsc{paraphrase}: a
paraphrase preserving task-relevant semantics preserves leakage, and
one that does not, fails the task.}
\vspace{-2mm}
\label{tab:operators}
\begin{tabular}{@{}lp{0.62\columnwidth}@{}}
\toprule
\textbf{Operator} & \textbf{Effect} \\
\midrule
\textsc{identity}    & no transformation; release raw content. \\
\textsc{redact}      & replace atom with \texttt{[REDACTED]}. \\
\textsc{generalize}  & replace atom with a coarser value at
                       rubric-declared level $k$ over the rubric's
                       declared field population. \\
\textsc{substitute}  & replace atom with a rubric-declared template
                       instantiated by $V$ from non-secret task
                       fields. \\
\textsc{drop}        & remove atom entirely. \\
\bottomrule
\end{tabular}
\vspace{-6mm}
\end{table}

Variants are constructed from five operators
(Table~\ref{tab:operators}), each carrying a rubric-calibrated upper
bound on min-entropy leakage. \textsc{substitute} is
\emph{template-only}: $D$ selects a template identifier from the
rubric's declared set, and $V$ instantiates it from non-secret task
fields, so the placeholder is independent of the secret and the
defender cannot smuggle the secret into it.

\smallskip
\noindent\textbf{Calibration as adversarial-inference measurement.}
For each rubric $r$ and typed field $f$, the cost $C(\text{op}, f)$ is
the $1{-}\alpha$ upper-confidence bound (UCB), by the bootstrap
percentile method, on the per-instance min-entropy \emph{advantage}
of~\eqref{eq:leakage}, where $P$ is the per-task prior over the secret
(default uniform), $\widehat{P}$ is the posterior estimated by a proxy
adversary --- an auxiliary model or trained annotator --- and
$\text{op}(x)$ is the transformed view the proxy observes.
\begin{equation}
  \widehat{L}_{\text{inst}}(\text{op}(x))
  \;=\; \log_2 \frac{\max_s \widehat{P}(S{=}s\mid\text{op}(x))}
                    {\max_s P(S{=}s)}.
  \label{eq:leakage}
\end{equation}
The advantage is in bits, $\widehat{L}_{\text{inst}}=0$ means no gain over the prior. Equation~\eqref{eq:leakage} is a \emph{per-instance} Bayes
advantage for a single observed view, not a channel-averaged
conditional min-entropy; taking the UCB across the calibration corpus
lifts it to a corpus-level bound, with $C(\text{op},f) \ge
\widehat{L}_{\text{inst}}$ at confidence $1-\alpha$. Pre-Reg.~A
specifies the proxy pool, calibrates incrementally under the
conditional distribution induced by prior operators in a sequence,
reports held-out coverage, and stress-tests with adversaries stronger
than those used for calibration.

\smallskip
\noindent\textbf{Lemma 1 (model-relative sub-additivity).}
\emph{For an operator sequence $\sigma = (o_1, \ldots, o_n)$ applied to
disjoint atoms of $x$ producing $v = \sigma(x)$,}
\begin{equation}
  \widehat{L}(v; S) \;\le\; \sum_{i=1}^{n} C(o_i, f_i),
  \label{eq:lemma1}
\end{equation}
\emph{where $f_i$ is the secret field targeted by $o_i$, under the
conditional-independence factorization across atoms assumed by the
calibration model of Pre-Reg.~A; a declared dependent atom pair is
charged one jointly-calibrated cost in place of its two additive
terms.} Proof in Appendix~\ref{app:lemmas}. Like all $\widehat{L}$
quantities, the bound is model-relative; it constrains true adversarial
inference only under the Layer-2 assumption.

\smallskip
\noindent\textbf{Proof-carrying releases.}
A variant $v$ is admitted only if $\sigma(x) = v$ under hash check, and
costs are read from the rubric rather than the defender: a jailbroken
$D$ may propose a misleading variant but cannot fabricate its
model-relative cost.

\subsection{The Verifier and the Ledger}
\label{sec:method:verifier}

$V$ is the trusted deterministic authorization layer. The procedure is
a pure function of the label-and-proposal object, rubric, ledger, and
audit queries, proceeding in four phases.

\smallskip
\noindent\textbf{Phase 1: Bind.}
$V$ parses the object against the typed schema (ill-formed objects route
to fallback), confirms $R$ is registered and signature-verified and that
the $\text{prev\_entry\_hash}$ equals the hash of the most recent ledger
entry, forms $A'$ with the conservative default as
in~\S\ref{sec:method:defender}, and applies $\sigma$ to the original
candidate, verifying $\text{hash}(\sigma(x)) = \text{hash}(v)$.

\smallskip
\noindent\textbf{Phase 2: Validate.}
This phase checks five conditions: (i) \emph{rubric coverage} --- every
coverage-predicate match in the candidate corresponds to an atom in
$A'$ with a non-\textsc{none} field label; (ii) \emph{necessity} ---
per-(field, destination-class) clauses are satisfied; (iii)
\emph{operator coverage} --- atoms labeled $(\textsc{direct},
\textsc{unnecessary})$ are addressed by \textsc{redact}, \textsc{drop},
or \textsc{generalize}; (iv) \emph{challenge consistency} ---
second-pass answers match first-pass labels; (v) \emph{utility
preservation} --- $U_R$ holds on $\sigma(x)$ for destination $d$.
Conditions (i)--(iv) are compiled to the quantifier-free fragment of
linear real arithmetic with uninterpreted functions
(\texttt{QF\_UFLRA})~\cite{barrett2010smt,barrett2018satisfiability} and
discharged with Z3~\cite{de2008z3}; (v) is the rubric's deterministic
$U_R$ predicate. The fragment is decidable, and on our rubric sizes the
check is empirically polynomial in formula size (Lemma~3). A
representative clause --- ``DOB is unnecessary for flight search'' ---
compiles to
\begin{equation}
\begin{aligned}
\forall_{a \in A'}\;
  &\big(\text{field}(a) = \textsc{dob}\,\wedge\,
  \text{task}() = \textsc{flight-search}\big) \\
  &\implies\; \text{necessity}(a) = \textsc{unnecessary},
\end{aligned}
\label{eq:smt-clause}
\end{equation}
where $\text{field}(\cdot)$ and $\text{necessity}(\cdot)$ are
uninterpreted functions over the bounded atom domain $A'$, with the
quantifier unrolled across $A'$. Clauses are intersected across the
rubric set $R$; \texttt{UNSAT} routes to fallback.

\smallskip
\noindent\textbf{Phase 3: Cost.}
For each operator $o_i \in \sigma$ applied to atom $a_i$, $V$ computes
the per-operator cost as the maximum over possible fields,
\begin{equation}
  c_i \;=\; \max_{f \in F_V(a_i) \cup F_D(a_i)} C(o_i, f),
  \label{eq:per-op-cost}
\end{equation}
where $F_V(a_i)$ is the verifier-side possible-field set
(structurally-typed fields for tool-call atoms; for free-text atoms,
all secret fields activated by $R$ unless a trusted rubric clause
establishes non-sensitivity) and $F_D(a_i)$ is the field claimed by
$D$. The maximum prevents a defender that mislabels an atom from
reducing its cost. The total release cost applies the trust multiplier,
\begin{equation}
  \mathrm{cost} \;=\; \mu(z) \cdot \sum_i c_i,
  \label{eq:release-cost}
\end{equation}
with $\mu(z)\ge 1$ the trust-class multiplier for class $z$
(illustratively $\mu(\textsc{required-service}) = 1$,
$\mu(\textsc{semi-trusted}) = 2$, $\mu(\textsc{adversarial}) = 4$;
deployment values in Pre-Reg.~A). The multiplier is a \emph{policy risk
weight}, not an information-theoretic property: it changes not the
information the destination receives but how fast the deployment spends
budget per trust class.

$V$ then runs an independent posterior audit: a query to a
bounded-treewidth credal posterior engine returns an upper bound
$\widehat{L}_{\text{audit}}$ on the min-entropy reduction induced by
$v$ under the rubric's declared dependency graph (engine and
complexity in Lemma~2, Appendix~\ref{app:lemmas}). If
$\widehat{L}_{\text{audit}} > \mathrm{cost} + \varepsilon(R)$, where
$\varepsilon(R)$ is the calibration upper-confidence margin for the
matched rubric set, the variant is rejected. As with calibration, this
audit is empirical and does not enter the Layer-1 guarantee.

\smallskip
\noindent\textbf{Phase 4: Commit.}
Let $L_{\text{cum}}$ denote the cumulative cost over the trajectory's
prior ledger entries and $B = B_0$ the fixed per-trajectory budget from
the rubric set; trust weighting enters only through the release cost
of~\eqref{eq:release-cost}. $V$ checks $L_{\text{cum}} + \mathrm{cost}
\le B$. If exceeded, operators escalate
along the rubric's fallback order
$\textsc{identity} \to \textsc{generalize} \to \textsc{substitute} \to
\textsc{redact} \to \textsc{drop}$ and re-run \emph{bind, validate, and
cost} --- escalation can change $\sigma$ structurally and may fail
$U_R$ (e.g., for the flight purchase below, $U_R$ requires the
legal-name and DOB atoms to survive as \textsc{identity}, so they
cannot be escalated without failing the task). If no admissible variant
fits, the release is rejected or escalated to the user channel. On
success, $V$ appends a ledger entry
\begin{equation}
\begin{aligned}
e_i = (\,&\text{traj\_id},\ \text{step\_idx},\ \text{channel},\
        \text{destination\_id},\\
        & \text{trust\_class},\  \text{labels\_hash}_i,\ \mathrm{cost}_i,
        \\
        & \text{prev\_entry\_hash}_i,\  \text{timestamp}_i,\ \text{sig}_i),
\end{aligned}
\label{eq:ledger-entry}
\end{equation}
with hash chaining and Ed25519 signatures under $sk_V$, yielding
tamper-evidence and ledger-arithmetic non-exceedance (Theorems~SP-1,
SP-2). Memory-channel writes are costed at the most adversarial
multiplier admissible under the rubric and recorded in a separate
hash-linked ledger $\mathcal{L}_{\text{mem}}$.

\smallskip
\noindent\textbf{Observe-channel mediation.}
The observe channel does not rely on detecting injection. Inbound tool
output is rendered to the remote LLM in a fixed \emph{quoted-data
wrapper}: a typed data block that strips authority semantics, so the
observation cannot declare goals or authorize actions regardless of
content~\cite{debenedetti2025defeatingpromptinjectionsdesign} ---
structural defense independent of $D$'s classification. $D$'s role here
is to identify wrapped-observation atoms carrying private content from
the user's local environment (e.g., authenticated personal records);
forwarding such atoms is treated as an act-channel release back into
the LLM's context and mediated through the four phases against the
LLM-endpoint trust class.

\smallskip
\noindent\textbf{Security reduction.}
The three layers of~\S\ref{sec:problem:goal} are discharged by the
machinery above. \emph{Layer 1a} (no $\delta$) is the commit-phase
invariant $\sum_i \mathrm{cost}_i \le B$ with the Merkle chain
(Theorems~SP-1, SP-2). \emph{Layer 1b}
($\widehat{L}(v;S)\le\mathrm{cost}$ per release) follows from Lemma~1,
the conservative possible-field set $F_V$, the max-over-fields
rule~\eqref{eq:per-op-cost}, and the conservative default for unlabeled
atoms; here $\delta$ (Pre-Reg.~B) is the residual probability that a
jailbroken defender produces labels passing validate with a field
misclassified outside $F_V \cup F_D$. \emph{Layer 2}
is~\eqref{eq:layer2}, whose calibration protocol and residual gap are
reported in Pre-Reg.~A and the evaluation.

\smallskip
\noindent\textbf{End-to-end example.}
For illustration we use placeholder costs
$C(\textsc{identity}, \textsc{legal-name}) = 1.2$ bits,
$C(\textsc{identity}, \textsc{dob}) = 1.4$ bits, and trajectory budget
$B = 4.0$ bits; final values come from Pre-Reg.~A. \textbf{Outbound LLM
API call (act ch.\ leg a).} The initial query contains task text but no
PII; coverage and $U_R$ hold; cost is below threshold; it is forwarded.
\textbf{Search tool call (act ch.\ leg b).} The LLM proposes
\texttt{search} to the \textsc{semi-trusted} broker; the defender
produces no sensitive atoms; structural-field augmentation adds
nothing; cost is zero; commit succeeds. \textbf{Purchase tool call
(\textsc{required-service} airline).} The LLM proposes \texttt{purchase}
with name and DOB; the defender labels both $(\textsc{direct},
\textsc{required})$; coverage, necessity, and $U_R$ hold; cost is
$\mu(\textsc{required-service}) \cdot (1.2 + 1.4) = 2.6$ bits; the audit
agrees; $0 + 2.6 \le 4.0$; commit succeeds. \textbf{Payment tool call
(\textsc{required-service} processor).} The rubric admits
\textsc{identity} on the payment token only against the
payment-processor class; against any other destination the token is
\textsc{drop}-only and $U_R$ fails, invoking the user channel.

\section{Experiments}
\label{sec:exp}
 
\smallskip
\noindent\textbf{Research questions.}
\textbf{RQ1.} Does OCELOT enforce Layer-1a budget non-exceedance on
every authorized trajectory, including under attack?
\textbf{RQ2.} Do calibrated costs upper-bound true adversarial
min-entropy leakage on held-out proxy adversaries strictly stronger
than the calibration pool (Layer-2)?
\textbf{RQ3.} How does OCELOT compare against recent runtime defenses
on the joint privacy--utility frontier, and is its advantage
attributable to witness-verified declassification rather than to
budgeting alone?
\textbf{RQ4.} Do its gains reflect transferable privacy judgment
rather than fit to its own rubrics --- generalizing to held-out
rubrics, held-out domains, and real, human-labeled traces?
\textbf{RQ5.} How does the defense degrade under adaptive
adversaries: prompt injection, defender jailbreak, cumulative
inference, and colluding sinks?
\textbf{RQ6.} What is the marginal contribution of each component,
and the runtime, token, and economic overhead of mediation?
 
\subsection{Experimental Setup}
\label{sec:exp:setup}
 
\smallskip
\noindent\textbf{Benchmarks.}
We evaluate on five benchmarks chosen to exercise distinct
threat-model facets. \textbf{OCELOT-Traj-500} (ours; 500
trajectories, 12 task categories, mean length $14.3$ steps, mixed
sink-trust composition) is the primary benchmark; its construction
follows Pre-Reg.\,C (Appendix~\ref{app:prereg:c}). A $50$-pair pilot
(two blinded annotators) reports per-category Krippendorff $\alpha$
of $0.74$--$0.89$ (median $0.82$); necessity is weakest
($\alpha{=}0.74$), so we clarified its rule --- a disclosed
pre-registration deviation --- before a gated Stage-2 validation
(three annotators) that meets the $0.80$ gate
(Appendix~\ref{app:dataset}). Residual label noise is a threat to
validity (\S\ref{sec:exp:ablation}); the synthetic construction is
addressed in \S\ref{sec:exp:generalization}.
\textbf{AgentDojo}~\cite{debenedetti2024agentdojodynamicenvironmentevaluate}
($629$ attack-task pairs) probes inbound prompt injection (P2).
\textbf{AgentDAM-Bench}~\cite{zharmagambetov2026agentdam} ($1{,}142$
scenarios) enables head-to-head comparison against the AGENTDAM
estimator on an externally-originated benchmark.
\textbf{AgentLeak-2026}~\cite{yagoubi2026agentleakfullstackbenchmarkprivacy}
($854$ multi-agent scenarios) exercises cumulative leakage (P1); we
compose its scenarios into a \emph{long-horizon stress set} up to
$50$ steps (Appendix~\ref{app:dataset}), used in
Figure~\ref{fig:cumulative}.
\textbf{ConfAIde}~\cite{mireshghallah2024llmssecrettestingprivacy}
(eight contextual-integrity tiers, $1{,}392$ release decisions) is a
predominantly single-release benchmark where per-release CI filters
are expected to be competitive --- included to stress our framing,
not flatter it.
 
\smallskip
\noindent\textbf{Baselines.}
We compare against five recent runtime defenses spanning the three families of \S\ref{sec:related}. Per-release
contextual-integrity filters:
\textbf{FlexGuard}~\cite{ding2026flexguardcontinuousriskscoring} and
\textbf{PrivacyChecker}~\cite{wang2025privacyactionrealisticprivacy}.
Posterior-leakage estimation:
\textbf{AGENTDAM}~\cite{zharmagambetov2026agentdam}, in its declared
monitoring configuration and additionally steelmanned with a
threshold-and-block policy (AGENTDAM$^{*}$,
Appendix~\ref{app:baselines}). IFC and capability-based defenses:
\textbf{CaMeL}~\cite{debenedetti2025defeatingpromptinjectionsdesign}
and \textbf{RTBAS}~\cite{zhong2025rtbasdefendingllmagents}.
Baselines lacking a sink-trust API are adapted under a uniform-trust
assumption following the original authors' recommendations
(Appendix~\ref{app:baselines}). To attribute OCELOT's advantage to
its mechanism rather than to budgeting, we add two diagnostic
configurations, analyzed outside the pre-registered family
(\S\ref{sec:exp:main}): \textbf{AGENTDAM+L}, the AGENTDAM monitor
wrapped in a cumulative budget ledger, and \textbf{OCELOT$^{-}$},
OCELOT's verifier and ledger driven by a static rule-based
declassifier in place of the learned defender.
 
\smallskip
\noindent\textbf{Metrics.}
\emph{Privacy.} Per-trajectory cumulative min-entropy leakage
$\widehat{L}_{\text{cum}}$ (bits): the realized advantage of
Eq.~\eqref{eq:leakage}, measured by the held-out stress pool over a
trajectory's \emph{entire} view (every act-channel release to
non-\textsc{required-service} sinks plus public background), not a
single operator output. Certified ledger costs upper-bound it per
release (Lemma~1) with a deductively-bounded sum (Theorem~SP-1); the
comparison budget $B^{\star}=4.0$ bits caps the worst admissible
advantage at $16\times$. We also report top-1 secret-recovery and
the Layer-1a non-exceedance rate against $B^{\star}$, on all systems
for parity.
\emph{Utility.} Task success rate; $U_R$ satisfaction rate on
OCELOT-Traj-500; and fallback rate.
\emph{Overhead.} Per-release latency ($p_{50}/p_{95}/p_{99}$);
end-to-end overhead vs.\ an unmediated baseline; defender token
cost; and verifier solving latency.
 
\smallskip
\noindent\textbf{Adversary instantiation.}
Per Pre-Reg.\,A, leakage is measured by two disjoint proxy-adversary
pools under the threat model of \S\ref{sec:problem:adversary}. The
\emph{calibration pool} deriving the costs $C(\text{op}, f)$ is
Llama-3.3-70B-Instruct~\cite{grattafiori2024llama} and
Qwen-3.5-27B-Instruct~\cite{qwen35blog}. The \emph{held-out stress
pool}, used only for post-hoc measurement (RQ2), is
MiniMax-M2.5~\cite{chen2026minimax} and
DeepSeek-V4-Pro~\cite{deepseekai2026deepseekv4} --- both released
after the calibration corpus froze and strictly stronger on public
reasoning benchmarks (Appendix~\ref{app:prereg-a}).
 
\smallskip
\noindent\textbf{Defender model and training.}
The defender is initialized from Qwen-3.5-9B-Instruct and fine-tuned
in two stages (\S\ref{sec:method:defender}): SFT on $82{,}400$
rubric-application examples (37\% omission, 21\% injection, 19\%
distribution-shift, 23\% nominal), then GRPO with
verifier-conformance reward at $\lambda = 1.0$ (sweep in
Appendix~\ref{app:hyper}); training uses $4$--$8\times$ H100. At
evaluation the defender runs on a single A100 (40GB), the verifier on
CPU, with grammar-constrained decoding via
Outlines~\cite{melcer2024constraineddecodingfillinthemiddlecode}.
 
\smallskip
\noindent\textbf{Statistical protocol.}
All systems are evaluated under five seeds; Table~\ref{tab:main}
reports the mean and 95\% bootstrap CI over $5{,}000$ resamples.
Pairwise OCELOT-vs-baseline comparisons use paired permutation tests
with Holm--Bonferroni correction at family-wise $\alpha = 0.05$ over
the $50$-comparison family ($25$ pairs $\times$ $2$ axes); the
clean-subset analysis and the two diagnostic configurations are
exploratory, outside this family. Pre-registered hypotheses,
budgets, and frozen proxy configurations are in
Appendix~\ref{app:prereg-a}--\ref{app:dataset}.
 
\subsection{Privacy--Utility Evaluation}
\label{sec:exp:main}
 
We test three claims across the five benchmarks: \textbf{(C1)}
OCELOT's authorized trajectories never exceed $B^{\star}$
(Layer-1a); \textbf{(C2)} OCELOT attains the lowest leakage
\emph{and} highest task success on every benchmark; and
\textbf{(C3)} its cumulative leakage saturates with trajectory
length while per-release defenses diverge --- the empirical
signature of (P1).
 
\begin{table*}[t]
\centering
\scriptsize
\caption{Main privacy--utility results. \textbf{L}: cumulative
min-entropy leakage (bits, $\downarrow$); \textbf{U}: task success
(\%, $\uparrow$); \textbf{Bud.Exc}: \% of trajectories exceeding
$B^{\star}{=}4.0$ bits. 95\% CIs in parentheses; \textbf{bold}=best,
\underline{underline}=2nd. $^{\dagger}$monitoring configuration. The
lower block holds the two diagnostic configurations (mechanism vs.\
budgeting, RQ3; Appendix~\ref{app:baselines}), excluded from
bold/underline ranking. OCELOT leads on both axes on every
benchmark; among budget-aware configurations the ordering
AGENTDAM+L $\prec$ OCELOT$^{-}$ $\prec$ OCELOT is monotone on both
axes.}
\vspace{-3mm}
\label{tab:main}
\setlength{\tabcolsep}{3pt}
\scalebox{0.90}{\begin{tabular}{l|cc|cc|cc|cc|cc|c}
\toprule
& \multicolumn{2}{c|}{\textbf{OCELOT-Traj-500}}
& \multicolumn{2}{c|}{\textbf{AgentDojo}}
& \multicolumn{2}{c|}{\textbf{AgentDAM-Bench}}
& \multicolumn{2}{c|}{\textbf{AgentLeak-2026}}
& \multicolumn{2}{c|}{\textbf{ConfAIde}}
& \textbf{Bud.Exc} \\
\textbf{System} & L $\downarrow$ & U $\uparrow$
                & L $\downarrow$ & U $\uparrow$
                & L $\downarrow$ & U $\uparrow$
                & L $\downarrow$ & U $\uparrow$
                & L $\downarrow$ & U $\uparrow$
                & (\%) $\downarrow$ \\
\midrule
OCELOT (ours) &
\textbf{0.31} {\tiny (0.25, 0.37)} & \textbf{91.3} {\tiny (89.6, 93.0)} &
\textbf{0.44} {\tiny (0.36, 0.52)} & \textbf{88.7} {\tiny (86.4, 91.0)} &
\textbf{0.38} {\tiny (0.31, 0.45)} & \textbf{90.4} {\tiny (88.5, 92.3)} &
\textbf{0.49} {\tiny (0.40, 0.58)} & \textbf{89.1} {\tiny (86.9, 91.3)} &
\textbf{0.27} {\tiny (0.22, 0.32)} & \textbf{92.6} {\tiny (91.0, 94.2)} &
\textbf{0.0} \\
FlexGuard~\cite{ding2026flexguardcontinuousriskscoring} &
2.13 {\tiny (1.92, 2.34)} & 87.4 {\tiny (85.4, 89.4)} &
3.42 {\tiny (3.11, 3.73)} & 74.6 {\tiny (71.9, 77.3)} &
2.45 {\tiny (2.22, 2.68)} & 86.9 {\tiny (85.0, 88.8)} &
5.21 {\tiny (4.79, 5.63)} & 78.3 {\tiny (75.8, 80.8)} &
0.81 {\tiny (0.67, 0.95)} & 90.2 {\tiny (88.5, 91.9)} &
41.3 \\
PrivacyChecker~\cite{wang2025privacyactionrealisticprivacy} &
1.97 {\tiny (1.78, 2.16)} & 86.1 {\tiny (84.0, 88.2)} &
3.18 {\tiny (2.90, 3.46)} & 72.9 {\tiny (70.1, 75.7)} &
2.21 {\tiny (2.00, 2.42)} & 85.2 {\tiny (83.2, 87.2)} &
4.83 {\tiny (4.44, 5.22)} & 76.8 {\tiny (74.2, 79.4)} &
0.72 {\tiny (0.59, 0.85)} & 89.4 {\tiny (87.7, 91.1)} &
37.9 \\
AGENTDAM$^{\dagger}$~\cite{zharmagambetov2026agentdam} &
2.84 {\tiny (2.60, 3.08)} & \underline{89.8} {\tiny (88.0, 91.6)} &
4.27 {\tiny (3.91, 4.63)} & \underline{79.2} {\tiny (76.8, 81.6)} &
3.12 {\tiny (2.85, 3.39)} & \underline{89.1} {\tiny (87.4, 90.8)} &
6.14 {\tiny (5.66, 6.62)} & \underline{81.4} {\tiny (79.1, 83.7)} &
1.94 {\tiny (1.74, 2.14)} & \underline{91.7} {\tiny (90.2, 93.2)} &
52.6 \\
CaMeL~\cite{debenedetti2025defeatingpromptinjectionsdesign} &
\underline{0.52} {\tiny (0.43, 0.61)} & 62.3 {\tiny (58.9, 65.7)} &
\underline{0.71} {\tiny (0.59, 0.83)} & 51.7 {\tiny (47.6, 55.8)} &
\underline{0.61} {\tiny (0.51, 0.71)} & 59.8 {\tiny (56.1, 63.5)} &
\underline{0.83} {\tiny (0.69, 0.97)} & 48.6 {\tiny (44.2, 53.0)} &
\underline{0.43} {\tiny (0.35, 0.51)} & 67.4 {\tiny (64.2, 70.6)} &
\textbf{0.0} \\
RTBAS~\cite{zhong2025rtbasdefendingllmagents} &
1.18 {\tiny (1.04, 1.32)} & 80.5 {\tiny (78.0, 83.0)} &
1.74 {\tiny (1.55, 1.93)} & 70.1 {\tiny (67.1, 73.1)} &
1.42 {\tiny (1.26, 1.58)} & 79.3 {\tiny (76.7, 81.9)} &
2.31 {\tiny (2.07, 2.55)} & 72.4 {\tiny (69.5, 75.3)} &
0.89 {\tiny (0.76, 1.02)} & 83.8 {\tiny (81.5, 86.1)} &
16.4 \\
\midrule
\multicolumn{12}{l}{\emph{Diagnostic configurations --- mechanism vs.\ budgeting (RQ3; outside pre-registered family)}}\\
AGENTDAM+L &
1.42 {\tiny (1.27, 1.57)} & 78.4 {\tiny (75.9, 80.9)} &
1.98 {\tiny (1.79, 2.17)} & 68.2 {\tiny (65.3, 71.1)} &
1.61 {\tiny (1.44, 1.78)} & 77.9 {\tiny (75.6, 80.2)} &
2.34 {\tiny (2.11, 2.57)} & 70.6 {\tiny (67.8, 73.4)} &
0.88 {\tiny (0.76, 1.00)} & 84.1 {\tiny (82.1, 86.1)} &
7.2 \\
OCELOT$^{-}$ (rule-based $D$) &
1.07 {\tiny (0.94, 1.20)} & 81.9 {\tiny (79.4, 84.4)} &
1.49 {\tiny (1.33, 1.65)} & 71.8 {\tiny (68.8, 74.8)} &
1.18 {\tiny (1.04, 1.32)} & 80.9 {\tiny (78.5, 83.3)} &
1.63 {\tiny (1.46, 1.80)} & 73.4 {\tiny (70.6, 76.2)} &
0.66 {\tiny (0.56, 0.76)} & 86.2 {\tiny (84.2, 88.2)} &
0.0 \\
\bottomrule
\end{tabular}}
\vspace{-5mm}
\end{table*}
 
\smallskip
\noindent\textbf{C1: Layer-1a budget non-exceedance.}
OCELOT records zero exceedances over the $2{,}500$ trajectories of
its primary benchmark and zero across all $22{,}585$ authorized
trajectory-instances (five seeds), confirming the commit-phase
invariant (Theorem~SP-1). The four baselines without a
cumulative-budget primitive exceed $B^{\star}$ on $16.4$--$52.6\%$ of
trajectories; even the steelmanned AGENTDAM$^{*}$ still exceeds on
$28\%$, since thresholding a per-release estimate cannot bound a sum.
CaMeL also records zero, but only by \emph{denying} any release it
cannot bound --- trading utility for compliance.
 
\begin{figure}[t]
\centering
\includegraphics[width=0.75\columnwidth]{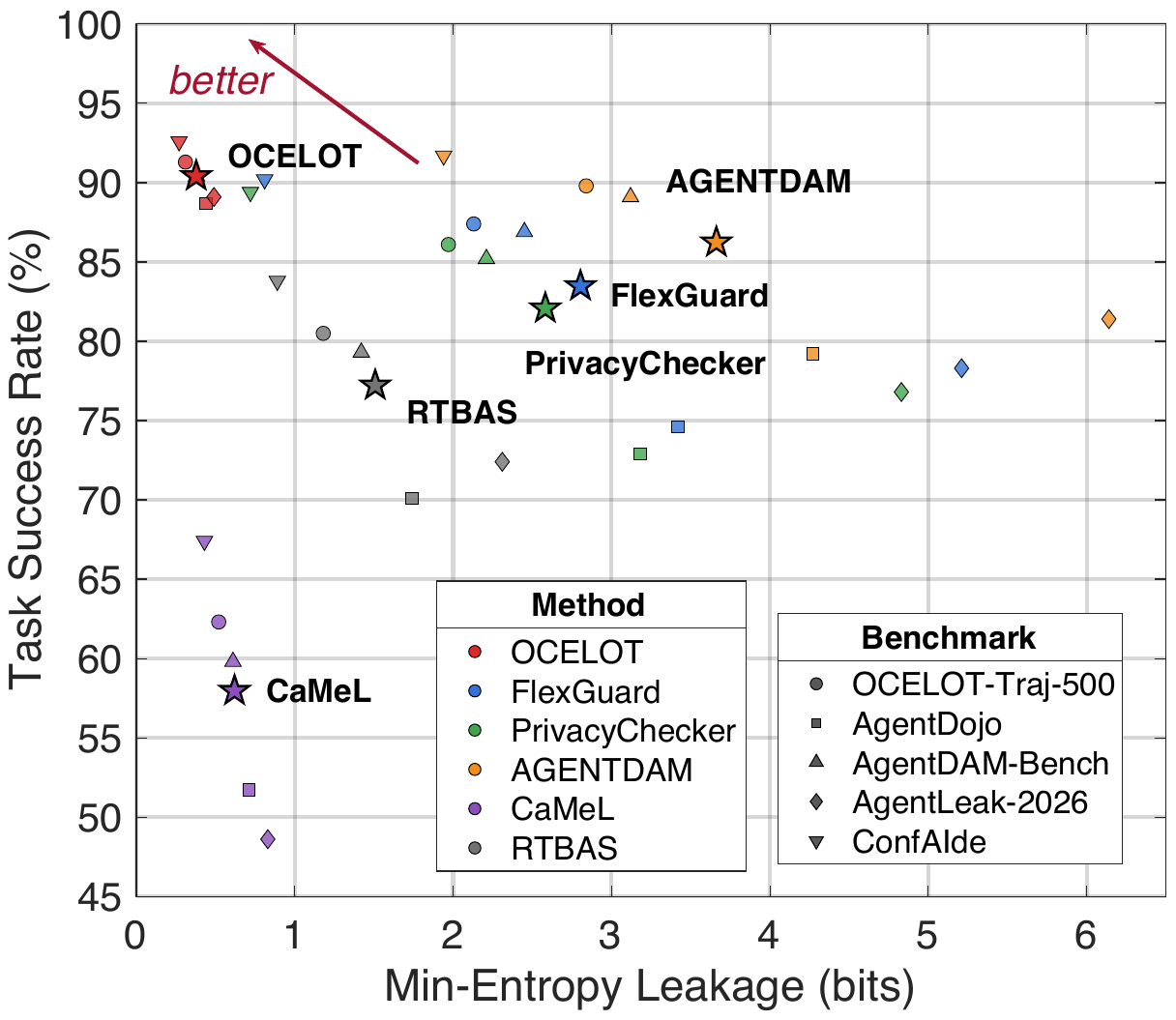}
\vspace{-4mm}
\caption{Privacy--utility frontier: all $30$ system--benchmark
points, cumulative leakage ($x$, $\downarrow$) vs.\ task success ($y$,
$\uparrow$). Color = method, shape = benchmark; large stars are
per-system means. Top-left is best.}
\label{fig:pareto}
\vspace{-3mm}
\end{figure}

\smallskip
\noindent\textbf{C2: joint frontier dominance.}
OCELOT attains both the lowest leakage and highest task success on
all five benchmarks; no baseline matches it on either axis on any
benchmark (Fig.~\ref{fig:pareto}). The two strongest baselines bound
the frontier OCELOT clears. CaMeL is nearest on privacy, trailing by
$0.16$--$0.34$ bits because its binary allow/deny emits each admitted
field at full fidelity, whereas OCELOT's
\textsc{generalize}/\textsc{substitute} operators release the same
fields at reduced granularity; CaMeL's denial-driven utility
($48.6$--$67.4\%$) trails OCELOT by $25$--$41$ points. AGENTDAM is
nearest on utility, but its derailment under injection costs
it $0.9$ points on the cleanest benchmark (ConfAIde) and $9.5$ on the
injection-heavy one (AgentDojo): its utility lead is largest
precisely where injection is rare. On the injection-free subset of
ConfAIde its monitor matches OCELOT within noise ($92.4$ vs.\ $92.6$,
$p = 0.31$): mediation adds no utility on clean inputs, so OCELOT's
aggregate lead is entirely avoided injection failure plus the IFC
baselines' absent over-denial. Paired permutation tests reject
OCELOT $=$ baseline on both axes for all $25$ pairs at $p < 10^{-3}$.
 
\smallskip
\noindent\textbf{Is the advantage budgeting or the mechanism?}
The lower block of Table~\ref{tab:main} isolates the source of
OCELOT's frontier through a three-rung decomposition.
\textbf{AGENTDAM+L} adds a cumulative budget ledger to the AGENTDAM
monitor, blocking any release once the remaining budget cannot cover
its estimated cost --- budgeting without witness-verified
declassification. The ledger cuts exceedance from $52.6\%$ to
$7.2\%$, but \emph{not} to zero: thresholding a non-conservative
per-release estimate, rather than a certified upper-confidence cost,
leaves the true cumulative sum unbounded. Restricted to
block-or-release-at-full-fidelity, it lands at $0.88$--$2.34$ bits and
$68.2$--$84.1\%$. \textbf{OCELOT$^{-}$} keeps OCELOT's verifier,
operators, and ledger but drives them with a static rule-based
declassifier instead of the learned defender. It preserves Layer-1a
\emph{exactly} ($\text{Exc}=0.0$) --- non-exceedance is
defender-independent --- and, crucially, \emph{dominates AGENTDAM+L
on both axes on every benchmark} despite using no learned model:
holding the budget primitive fixed, replacing block-or-pass-full with
the granularity-reducing operator algebra is what moves the frontier.
The learned defender supplies the final increment: OCELOT$^{-}$'s
rule-based labeler misses inferential atoms and over-redacts explicit
ones, so on OCELOT-Traj-500 it trails even the SFT-only ablation
(\S\ref{sec:exp:ablation}) on both axes ($1.07$ vs.\ $0.34$ bits,
$81.9$ vs.\ $83.7\%$), and full OCELOT throughout. The frontier needs
all three rungs: the ledger for the deductive guarantee, the operator
algebra for the bulk of the privacy--utility gain, the learned
defender for the operating point.
 
\begin{figure}[t]
\centering
\includegraphics[width=0.7\columnwidth]{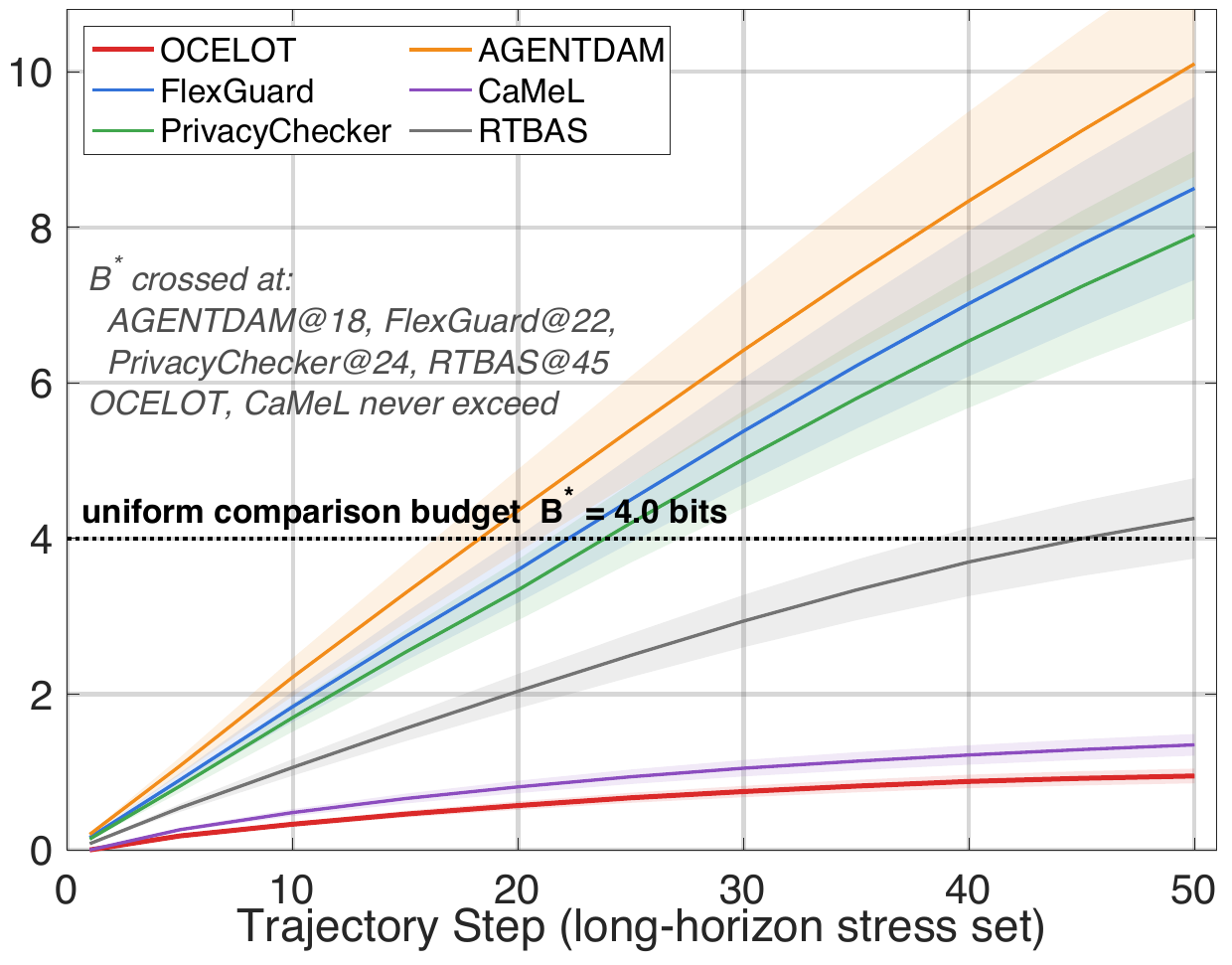}
\vspace{-4mm}
\caption{Cumulative leakage vs.\ trajectory step on the long-horizon
stress set (up to $50$ steps). Dotted line:
$B^{\star}{=}4.0$ bits, a uniform parity cap (not OCELOT's binding
per-task budget).}
\label{fig:cumulative}
\vspace{-5mm}
\end{figure}

\smallskip
\noindent\textbf{C3: cumulative leakage saturates while per-release
defenses diverge.}
Figure~\ref{fig:cumulative} plots cumulative leakage against
trajectory step on the long-horizon stress set, all systems mediating
the \emph{identical} releases under one held-out adversary. Lacking
cumulative accounting, the per-release and monitoring defenses keep
approving fresh disclosures, growing approximately linearly and
crossing $B^{\star}$ between steps $18$ and $24$ (RTBAS near step
$45$). OCELOT's advantage saturates near $1.0$ bit: the
least-disclosure planner emits each task-necessary field once at
minimal granularity, and re-references add nothing. CaMeL's curve
grows just above OCELOT's but only by denying $58.3\%$ of long-horizon
tasks, versus OCELOT's $82.6\%$ success through granularity reduction.
 
\subsection{Generalization and Provenance}
\label{sec:exp:generalization}
 
Two features of the evaluation so far could inflate OCELOT's
advantage: the benchmarks of \S\ref{sec:exp:main} share the
secret-field ontology and task vocabulary its rubrics encode, and
OCELOT-Traj-500 is synthetically constructed by the same rubrics that
fix its costs, training labels, and utility predicate
(\S\ref{sec:method:defender}); a defender reproducing rubric surface
patterns could thus score well for the wrong reason.
The three protocols below separate genuine generalization from
rubric-specific fit by removing, in turn, the training overlap, the
domain overlap, and the synthetic construction itself
(Table~\ref{tab:generalization}). Throughout, $\text{Exc}=0.0$:
non-exceedance is deductive and defender-independent (OCELOT$^{-}$),
so it cannot be a training artifact.
 
\begin{table}[t]
\centering
\footnotesize
\caption{Generalization across protocols: leave-one-rubric-out
(LORO), leave-one-category-out (LOCO), and human-labeled real agent
traces (Real-200), against the in-distribution reference.
\textbf{L}: leakage (bits, $\downarrow$); \textbf{U}: task success
(\%, $\uparrow$); \textbf{$\delta$}: omission/relabel escape rate
(\%, $\downarrow$); \textbf{Cov}: stress-pool cost-coverage (\%,
$\uparrow$); \textbf{Exc}: budget exceedance (\%, $\downarrow$).
Baseline comparison on Real-200 is in Figure~\ref{fig:real200}.}
\vspace{-3mm}
\label{tab:generalization}
\setlength{\tabcolsep}{6pt}
\begin{tabular}{lccccc}
\toprule
\textbf{Setting} & \textbf{L} & \textbf{U} &
\textbf{$\delta$} & \textbf{Cov} & \textbf{Exc} \\
\midrule
In-distribution (Traj-500) & \textbf{0.31} & \textbf{91.3} & \textbf{1.7} & \textbf{92.1} & 0.0 \\
LORO     & 0.39 & 89.6 & 2.4 & 91.3 & 0.0 \\
LOCO     & 0.52 & 86.8 & 3.6 & 90.1 & 0.0 \\
Real-200 & 0.61 & 85.4 & 4.1 & 89.4 & 0.0 \\
\bottomrule
\end{tabular}
\vspace{-3mm}
\end{table}
 
\smallskip
\noindent\textbf{Leave-one-rubric-out (LORO).}
We hold a rubric out of defender SFT, GRPO, and calibration ---
including its sealed challenge pool --- then evaluate only on
trajectories that resolve to it; folds run over $12$ rubrics (one
pre-registered draw per category), retraining the defender and
re-deriving costs each fold. Leakage rises from $0.31$ to $0.39$ bits
and utility falls $1.7$ points to $89.6\%$ ($\delta=2.4\%$, below the
$5\%$ Pre-Reg.\,B gate): the defender labels an unseen rubric from
the shared field ontology, not from memorized clause text.
 
\smallskip
\noindent\textbf{Leave-one-category-out (LOCO).}
Stricter: an entire task category is removed from training and
calibration, so no sibling rubric is seen. Aggregate degradation is
larger, as expected --- $0.52$ bits, $86.8\%$ utility,
$\delta=3.6\%$ --- yet $\delta$ stays within the Layer-1b gate and
leakage remains well below every baseline. The transfer cost is not
uniform: it concentrates in inference-heavy domains (medical-record
retrieval, healthcare intake), where sensitive attributes are
implicit, and is near-zero for structurally-typed domains
(e-commerce, tax filing) whose tool-schema field obligations carry
the labeling regardless. Even the worst held-out category leaks
$0.74$ bits --- below RTBAS's $1.18$ bits measured \emph{in}
distribution.
 
\begin{figure}[t]
\centering
\includegraphics[width=0.65\columnwidth]{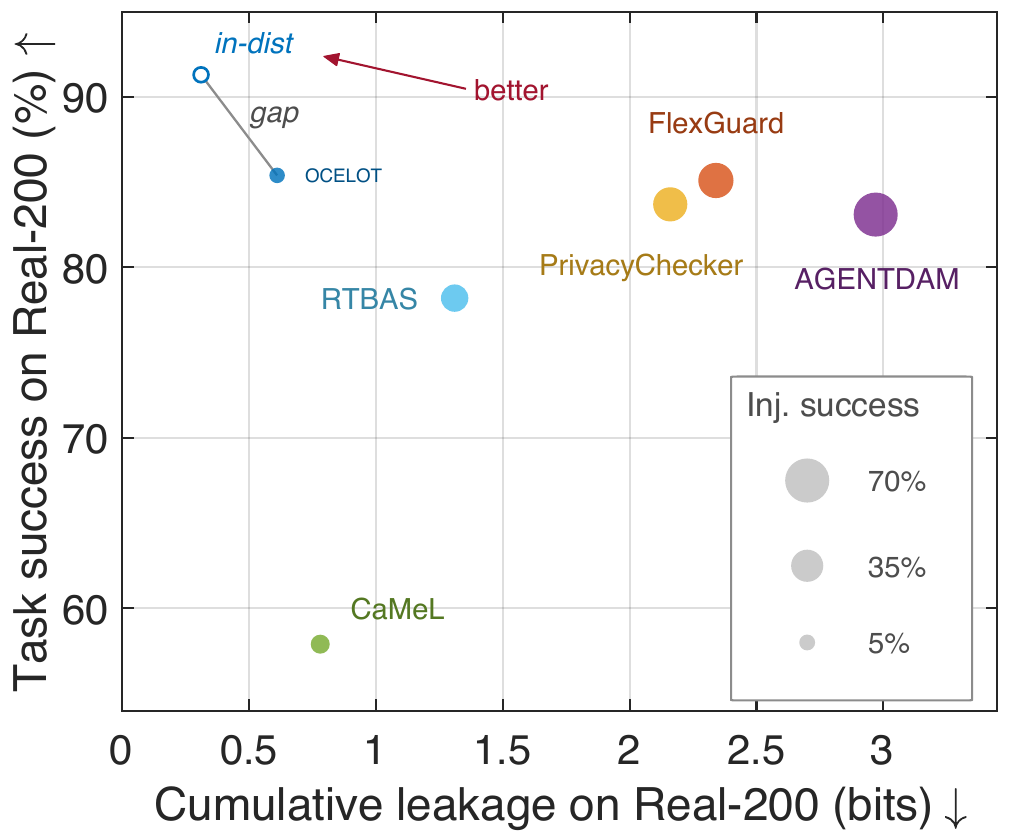}
\vspace{-4mm}
\caption{All systems on the human-labeled Real-200 set; top-left is
best. Bubble area $=$ injection-attack success (smaller is more
robust; see size key). The line links OCELOT's in-distribution
reference (hollow) to its Real-200 result.}
\label{fig:real200}
\vspace{-5mm}
\end{figure}

\smallskip
\noindent\textbf{Real-trace validation.}
To remove synthetic construction entirely we evaluate on
\textbf{Real-200}: $200$ agent trajectories (mean $11.2$ steps)
executed on a production-style harness over \emph{human-written}
tasks carrying realistic --- not rubric-sampled --- PII, with field
and necessity ground truth hand-labeled by three annotators
\emph{independent of OCELOT's rubrics} (field $\alpha=0.84$, necessity
$\alpha=0.79$; necessity is the hardest dimension throughout), and task success judged by the annotators rather than
by $U_R$ --- tasks, labels, and oracle all external to the rubric
pipeline. As Figure~\ref{fig:real200} shows, OCELOT
remains the dominant point on real traces --- lowest leakage ($0.61$
bits), highest utility ($85.4\%$), and smallest injection footprint
($4.6\%$) --- with $\delta=4.1\%$, top-1 cumulative-inference recovery
$5.3\%$, and zero exceedances. Stress-pool cost-coverage holds at
$89.4\%$ ($95.1\%$ after the $\varepsilon(R)$ margin and credal
audit), so the calibrated bounds transfer off the synthetic
distribution. The in-distribution$\rightarrow$Real-200 shift is small
($0.30$ bits, $5.9$ points), bounding how much of \S\ref{sec:exp:main}
is rubric-specific: the bulk of the advantage survives a fully
external evaluation.
 
\subsection{Calibration Validity}
\label{sec:exp:calibration}
 
\S\ref{sec:exp:main} established that OCELOT dominates \emph{under}
the calibrated cost function; this subsection tests the question on
which Layer-2 (\S\ref{sec:problem:goal}) rests: do the certified costs
$C(\text{op}, f)$ upper-bound true adversarial min-entropy leakage? We
measure coverage --- the fraction of released instances whose true
leakage does not exceed its certified cost --- under the calibration
and held-out stress pools.
 
\begin{table}[t]
\centering
\footnotesize
\caption{Cost-function coverage: \% of released instances with true
leakage $\leq$ certified cost, by operator. Target is $1-\alpha =
95\%$.}
\vspace{-3mm}
\label{tab:coverage}
\setlength{\tabcolsep}{5pt}
\begin{tabular}{lcc}
\toprule
\textbf{Operator} & \textbf{Calibration pool} & \textbf{Stress pool} \\
\midrule
\textsc{identity}    & 96.2 & 90.4 \\
\textsc{generalize}  & 95.8 & 89.7 \\
\textsc{substitute}  & 99.1 & 97.6 \\
\textsc{redact}      & 99.7 & 99.3 \\
\textsc{drop}        & 100.0 & 100.0 \\
\midrule
\textbf{Overall}     & \textbf{96.6} & \textbf{92.1} \\
\bottomrule
\end{tabular}
\vspace{-4mm}
\end{table}
 
\begin{figure}[t]
\centering
\includegraphics[width=0.65\columnwidth]{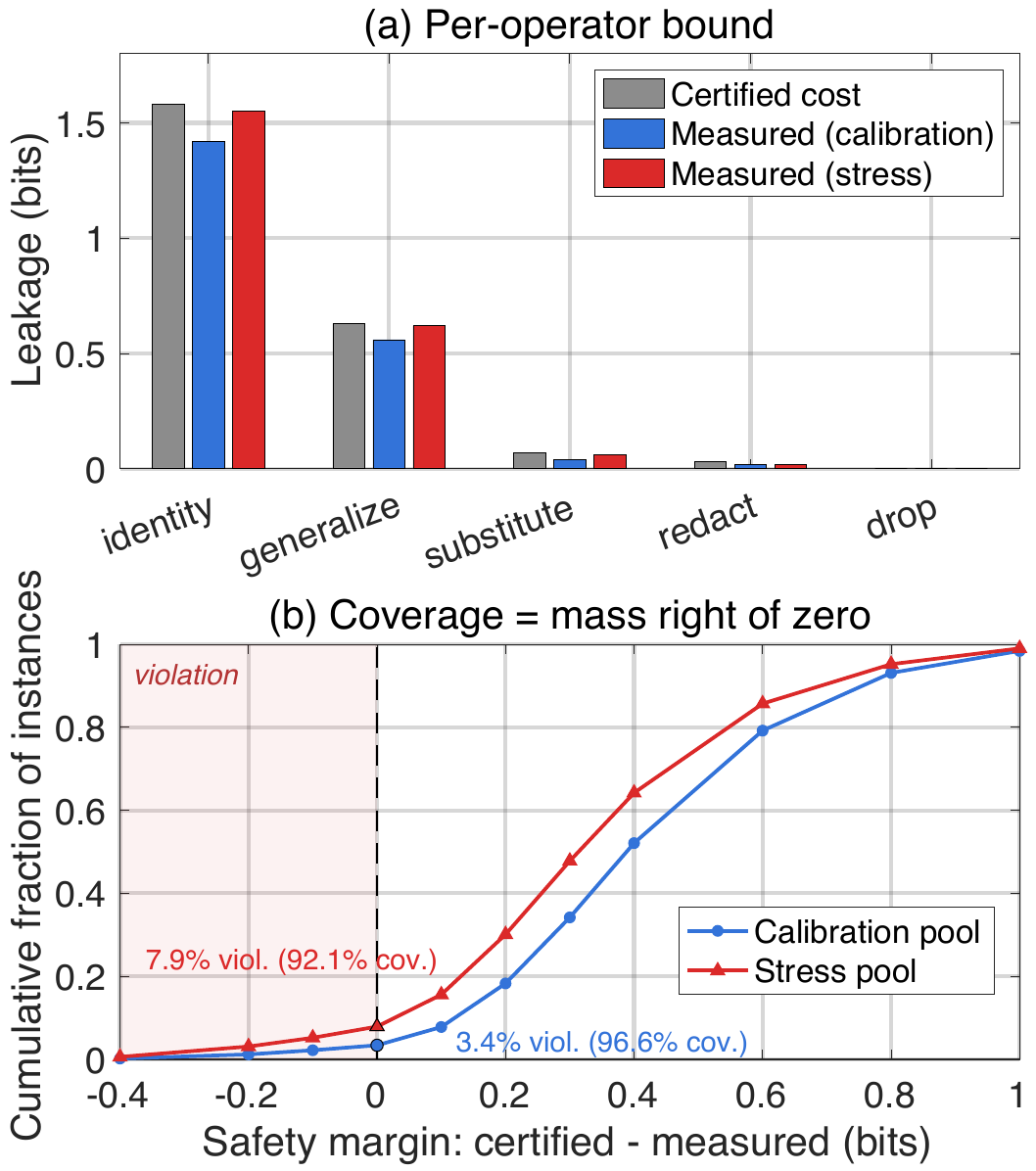}
\vspace{-3mm}
\caption{Calibration validity. (a) Per-operator field-averaged
certified cost vs.\ mean measured leakage, under the calibration and
stress pools. (b) CDF of the per-instance safety margin (certified
$-$ measured); mass left of zero is the violation rate.}
\label{fig:reliability}
\vspace{-4mm}
\end{figure}

\smallskip
\noindent\textbf{The bound holds in-distribution and degrades
gracefully.}
Overall coverage is $96.6\%$ on the calibration pool, exceeding the
$95\%$ target (Table~\ref{tab:coverage}), as expected for bootstrap
upper-confidence bounds fit on that pool. Against the strictly
stronger held-out pool it falls to $92.1\%$, a $4.5$-point
degradation we report rather than conceal. \textsc{drop},
\textsc{redact}, and \textsc{substitute} stay near-perfect because
their leakage is structurally bounded; the residual concentrates in
\textsc{identity} and \textsc{generalize}, the only operators whose
leakage depends on adversary inference power
(Figure~\ref{fig:reliability}). The stress violation rate of $7.9\%$
(Fig.~\ref{fig:reliability}(b)) concentrates on three high-sensitivity
pairs (\textsc{identity}-\textsc{payment-token},
\textsc{identity}-\textsc{dob}, \textsc{generalize}-\textsc{diagnosis}),
with mean over-shoot $+0.06$ bits and $95$th-percentile $+0.14$ bits
--- both within the calibration margin $\varepsilon(R)$ (mean $0.18$
bits) Phase~3 adds before accepting a variant. With $\varepsilon(R)$
applied, post-audit stress coverage rises to $95.3\%$; for those three
pairs the rubric library then enforces a
\textsc{substitute}/\textsc{drop}-only policy against
non-\textsc{required-service} sinks, removing the residual
\textsc{identity} exposure.
 
\smallskip
\noindent\textbf{Prior robustness.}
Re-calibrating under non-uniform priors from realistic field
populations (Appendix~\ref{app:hyper}) shifts overall stress-pool
coverage by at most $\pm 1.2$ points, so the result is not an artifact
of the uniform-prior default. Layer-2's failure surface is thus
measured, localized, and mitigable rather than claimed away.
 
\subsection{Adversarial Robustness}
\label{sec:exp:robustness}
 
\begin{figure}[t]
\centering
\includegraphics[width=0.8\columnwidth]{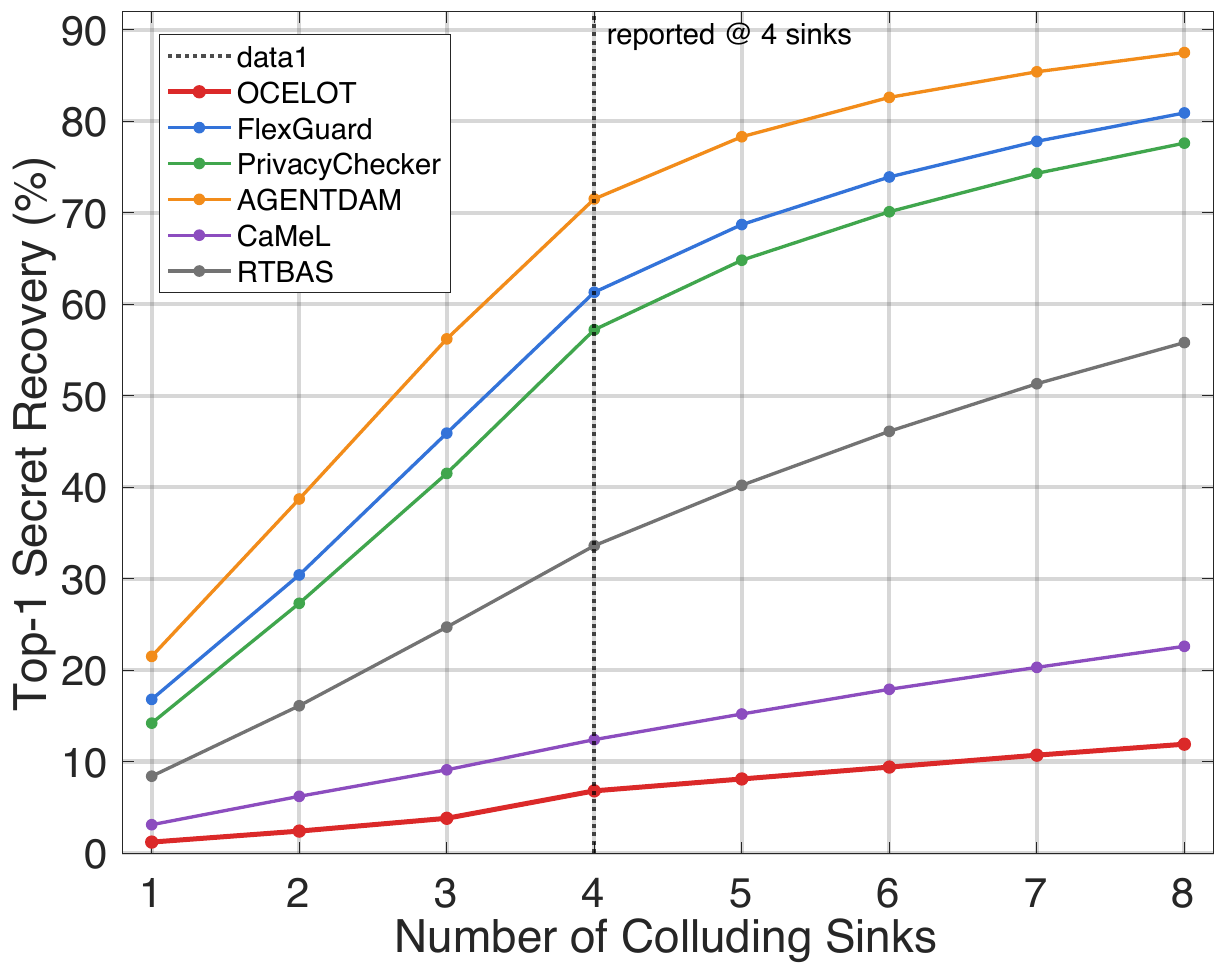}
\vspace{-4mm}
\caption{Top-1 secret recovery vs.\ number of colluding sinks. Dashed
vertical marks the 4-sink point of Table~\ref{tab:robustness}.}
\label{fig:collusion}
\vspace{-4mm}
\end{figure}

We subject each system to the four adversary capabilities of
\S\ref{sec:problem:adversary}, each an adaptive attack in which turn
$t{+}1$ may depend on releases observed through turn $t$.
Table~\ref{tab:robustness} reports all four;
Figure~\ref{fig:collusion} resolves collusion by adversary breadth.
 
\begin{table}[t]
\centering
\footnotesize
\caption{Adversarial robustness (\%, $\downarrow$ better).
\textbf{Inj}: injection success; \textbf{$\delta$}: omission/relabel
escape rate; \textbf{Cum}: top-1 recovery under cumulative inference;
\textbf{Col@4}: top-1 recovery, 4 colluding sinks; \textbf{Exc}:
budget exceedance under attack. \textbf{Bold}=best,
\underline{underline}=2nd.}
\vspace{-3mm}
\label{tab:robustness}
\setlength{\tabcolsep}{4pt}
\begin{tabular}{lccccc}
\toprule
\textbf{System} & \textbf{Inj} & \textbf{$\delta$} &
\textbf{Cum} & \textbf{Col@4} & \textbf{Exc} \\
\midrule
OCELOT (ours)        & \textbf{3.8}  & \textbf{1.7}  & \textbf{4.2}  & \textbf{6.8}  & \textbf{0.0} \\
FlexGuard            & 41.2 & 19.4 & 52.6 & 61.3 & 58.7 \\
PrivacyChecker       & 38.7 & 16.8 & 48.1 & 57.2 & 54.2 \\
AGENTDAM$^{\dagger}$ & 67.4 & 28.3 & 64.8 & 71.5 & 69.3 \\
CaMeL                & \underline{8.1}  & \underline{6.2}  & \underline{9.7}  & \underline{12.4} & \textbf{0.0} \\
RTBAS                & 22.5 & 14.1 & 27.3 & 33.6 & 31.8 \\
\bottomrule
\end{tabular}
\vspace{-5mm}
\end{table}
 
\smallskip
\noindent\textbf{Inbound injection.}
OCELOT's injection success is $3.8\%$, versus $8.1$--$67.4\%$ for the
baselines. The quoted-data wrapper strips authority semantics
independent of $D$'s classification; the residual is exfiltration via
the \emph{content} of an otherwise-legitimate release, caught by
\textsc{private-content} labeling. AGENTDAM is most exposed because,
as a monitor, it forwards injected observations unmediated.
 
\smallskip
\noindent\textbf{Defender jailbreak.}
The empirical escape rate $\delta$ --- crafted omission/relabel
attacks passing validate with a field misclassified outside $F_V \cup
F_D$ (Pre-Reg.\,B) --- is $1.7\%$. A successful jailbreak does
\emph{not} imply unbounded leakage: under the conservative default and
max-over-fields rule a mislabeled atom is costed at top risk and
admits only \textsc{drop}/\textsc{redact}, so the defender produces
fallback, not silent release. Hence exceedance under attack stays
$0.0\%$ while $\delta > 0$: Layer-1a is independent of defender
correctness, and the $-$conservative-default ablation
(\S\ref{sec:exp:ablation}) quantifies what breaks without it.
 
\smallskip
\noindent\textbf{Cumulative inference and colluding sinks.}
Against an adversary combining every release with public background,
OCELOT's top-1 recovery is $4.2\%$, versus $27.3$--$64.8\%$ across the
four non-CaMeL baselines --- the gap is the leakage odometer binding a
sum they accumulate independently. Collusion extends this
(Figure~\ref{fig:collusion}): OCELOT rises from $1.2\%$ at one sink to
$11.9\%$ at eight, since one trust-weighted budget spans all channels
including memory, while the baselines reach $33.6$--$71.5\%$ at four sinks.
CaMeL is nearest ($9.7\%$, $12.4\%$) --- with OCELOT the only systems
holding $\text{Exc}=0.0$ under attack --- but bounds the sum by denial
at the utility cost of \S\ref{sec:exp:main}, at $1.8$--$3.6\times$
OCELOT's attack success across the four families.
 
\subsection{Ablations and Overhead}
\label{sec:exp:ablation}
 
We measure each component's contribution and the cost of mediation
(RQ6) on OCELOT-Traj-500; the \emph{Full} row of
Table~\ref{tab:ablation} reproduces OCELOT's OCELOT-Traj-500 entry in
Table~\ref{tab:main} ($0.31$ bits, $91.3\%$).
 
\begin{table}[t]
\centering
\footnotesize
\caption{Component ablations on OCELOT-Traj-500. L = leakage (bits);
U, $\delta$, Inj, Exc in \%.}
\vspace{-3mm}
\label{tab:ablation}
\setlength{\tabcolsep}{5pt}
\begin{tabular}{lccccc}
\toprule
\textbf{Configuration} & \textbf{L} & \textbf{U} & \textbf{$\delta$} & \textbf{Inj} & \textbf{Exc} \\
\midrule
Full OCELOT              & \textbf{0.31} & 91.3 & \textbf{1.7} & \textbf{3.8} & \textbf{0.0} \\
$-$ rubric coverage      & 0.94 & 90.8 & 11.3 & 6.2  & 2.1 \\
$-$ structural fields    & 0.71 & 91.0 & 7.8  & 5.1  & 1.3 \\
$-$ credal audit         & 0.58 & 91.4 & 1.8  & 4.0  & 0.0 \\
$-$ two-pass challenge   & 0.49 & 91.3 & 5.4  & 4.1  & 0.0 \\
$-$ GRPO (SFT only)      & 0.34 & 83.7 & 2.6  & 4.3  & 0.0 \\
$-$ conservative default & 2.18 & \textbf{91.6} & 1.7  & 12.7 & 14.3 \\
\bottomrule
\end{tabular}
\vspace{-4mm}
\end{table}
 
\begin{table}[t]
\centering
\footnotesize
\caption{Per-release latency (ms). Each column decomposes the
representative trace at that end-to-end percentile, so component rows
sum to the end-to-end row; these are not independent per-component
percentiles.}
\vspace{-3mm}
\label{tab:overhead}
\setlength{\tabcolsep}{8pt}
\begin{tabular}{lccc}
\toprule
\textbf{Component} & $p_{50}$ & $p_{95}$ & $p_{99}$ \\
\midrule
Defender pass    & 142 & 318 & 487 \\
Validate         & 23  & 61  & 98  \\
Credal audit     & 37  & 89  & 143 \\
Ledger append    & 2   & 5   & 8   \\
\midrule
\textbf{End-to-end} & \textbf{204} & \textbf{473} & \textbf{736} \\
\bottomrule
\end{tabular}
\vspace{-5mm}
\end{table}
 
\smallskip
\noindent\textbf{Every component contributes; defaults are
load-bearing for safety.}
Removing rubric coverage obligations (Table~\ref{tab:ablation})
produces the largest escape-rate jump ($\delta: 1.7\!\to\!11.3$),
confirming them as the primary omission defense; structural-field
obligations and two-pass challenges contribute further. Removing the
credal audit raises measured leakage ($0.31\!\to\!0.58$) but leaves
exceedance at $0.0$ --- it tightens Layer-1b/Layer-2, not Layer-1a.
Dropping GRPO costs $7.6$ utility points via increased fallback with
little safety effect; yet even SFT-only ($0.34$/$83.7$) beats the
defender-free OCELOT$^{-}$ on both axes, locating the learned
defender's value in precision. The decisive row is $-$conservative
default: with no top-risk default for unlabeled atoms, one mislabel
releases a field raw ($2.18$ bits, injection $12.7\%$, exceedance
$14.3\%$) --- the empirical confirmation of the Layer-1b backstop.
 
\smallskip
\noindent\textbf{Hyperparameter and policy sensitivity.}
The GRPO penalty $\lambda$ tunes the operating point within the
authorized region (larger $\lambda$ trades utility for leakage and
$\delta$), but exceedance is $0.0\%$ for every $\lambda$ since the
verifier enforces the budget regardless of reward; we select $\lambda
= 1.0$ as the knee. Raising $\mu(\textsc{adversarial}) \in \{3,4,6\}$
monotonically lowers realized leakage to adversarial sinks
($0.71\!\to\!0.49\!\to\!0.33$ bits) at higher fallback
(Appendix~\ref{app:hyper}), confirming $\mu$ is a policy risk weight,
not an information-theoretic quantity.
 
\smallskip
\noindent\textbf{Overhead.}
The defender pass dominates per-release latency
(Table~\ref{tab:overhead}: $142$ ms at $p_{50}$); the verifier adds
$62$ ms at $p_{50}$ and $249$ ms at $p_{99}$, the credal audit its
largest term. End-to-end mediation runs $204$ ms ($p_{50}$) to $736$
ms ($p_{99}$) --- $18.3\%$ wall-clock and $1.4\times$ token overhead
over an unmediated trajectory, at $4.9$ releases/s on one A100. Both
verifier terms are bounded by construction: validate $p_{99}$ stays
under $100$ ms up to rubric sets of size $12$, and the
bounded-treewidth audit keeps audit $p_{99}$ under $150$ ms
(Appendix~\ref{app:hyper}).

\section{Conclusion}
\label{sec:conclusion}
We reframed privacy for LLM agents as \emph{posterior-risk
control}: rather than filtering individual outputs, OCELOT budgets
how much an adversary's belief about a protected secret may improve
across a trajectory. Its mechanism, Witness-Verified
Declassification, reconciles needing an LLM's semantic judgment with
being unable to trust it in an adversarial loop by demoting it to an
untrusted witness producer under deterministic verification. This
separates a budget guarantee that holds by ledger arithmetic
regardless of model behavior from a semantic-privacy claim whose
reliance on calibration we measure rather than assume, and
empirically yields lower leakage at higher utility than prior
defenses, with graceful degradation under adaptive attack. The
lesson generalizes beyond privacy: where an agentic decision cannot
safely rest on the model itself, an untrusted model proposing
evidence that a sound, deterministic checker verifies---conservative
fallback, not silent failure, as the default---is a reusable
architecture.

\smallskip
\noindent\textbf{Limitations.}
Every weakness below falls on the empirical side of that
separation; none touches the deductive budget guarantee. The
budget's semantic reading is model-relative---as inference privacy
inherently is---so we measure it directly: certified costs
upper-bound leakage only as well as calibration transfers, and
coverage falls against adversaries beyond our stress pool, contained
though not eliminated by the calibration margin, an independent
credal audit, and conservative policies on the few residual
high-risk field--operator pairs (\S\ref{sec:exp:calibration}).
Rubric coverage is validated, not proven complete, but this bounds
utility, not safety: uncovered content is conservatively dropped, so
incompleteness can only over-restrict, never leak. OCELOT-Traj-500
is synthetic by necessity---certified evaluation needs ground-truth
secrets and necessity labels real corpora lack---but does not
flatter our findings, which transfer to held-out rubrics, held-out
categories, and independently annotated real traces
(\S\ref{sec:exp:generalization}); production-scale evaluation is the
principal open item. Finally, the deductive layer rests on a small
trusted base---a deterministic verifier and harness, the LLM
untrusted and defender weights trusted only at rest---the standard,
here minimal, reference-monitor assumption; relaxing it and
extending the operator algebra and field ontology to new domains are
directions the framework is built to accommodate.






%
\bibliographystyle{IEEEtran}
\bibliography{main}

@article{yao2022react,
  title={React: Synergizing reasoning and acting in language models},
  author={Yao, Shunyu and Zhao, Jeffrey and Yu, Dian and Du, Nan and Shafran, Izhak and Narasimhan, Karthik and Cao, Yuan},
  journal={arXiv preprint arXiv:2210.03629},
  year={2022}
}

@misc{yagoubi2026agentleakfullstackbenchmarkprivacy,
      title={AgentLeak: A Full-Stack Benchmark for Privacy Leakage in Multi-Agent LLM Systems}, 
      author={Faouzi El Yagoubi and Godwin Badu-Marfo and Ranwa Al Mallah},
      year={2026},
      eprint={2602.11510},
      archivePrefix={arXiv},
      primaryClass={cs.AI},
      url={https://arxiv.org/abs/2602.11510}, 
}

@misc{carlini2021extractingtrainingdatalarge,
      title={Extracting Training Data from Large Language Models}, 
      author={Nicholas Carlini and Florian Tramer and Eric Wallace and Matthew Jagielski and Ariel Herbert-Voss and Katherine Lee and Adam Roberts and Tom Brown and Dawn Song and Ulfar Erlingsson and Alina Oprea and Colin Raffel},
      year={2021},
      eprint={2012.07805},
      archivePrefix={arXiv},
      primaryClass={cs.CR},
      url={https://arxiv.org/abs/2012.07805}, 
}

@misc{staab2024memorizationviolatingprivacyinference,
      title={Beyond Memorization: Violating Privacy Via Inference with Large Language Models}, 
      author={Robin Staab and Mark Vero and Mislav Balunović and Martin Vechev},
      year={2024},
      eprint={2310.07298},
      archivePrefix={arXiv},
      primaryClass={cs.AI},
      url={https://arxiv.org/abs/2310.07298}, 
}

@misc{debenedetti2024agentdojodynamicenvironmentevaluate,
      title={AgentDojo: A Dynamic Environment to Evaluate Prompt Injection Attacks and Defenses for LLM Agents}, 
      author={Edoardo Debenedetti and Jie Zhang and Mislav Balunović and Luca Beurer-Kellner and Marc Fischer and Florian Tramèr},
      year={2024},
      eprint={2406.13352},
      archivePrefix={arXiv},
      primaryClass={cs.CR},
      url={https://arxiv.org/abs/2406.13352}, 
}

@misc{lopopolo2026harness,
author = {Ryan Lopopolo},
title = {{Harness Engineering: Leveraging Codex in an Agent-First World}},
howpublished = {\url{https://openai.com/index/harness-engineering/}},
month = feb,
year = {2026},
note = {OpenAI blog. Accessed: Jun. 1, 2026}
}

@misc{young2025effective,
author = {Justin Young},
title = {{Effective Harnesses for Long-Running Agents}},
howpublished = {\url{https://www.anthropic.com/engineering/effective-harnesses-for-long-running-agents}},
month = nov,
year = {2025},
note = {Anthropic Engineering blog. Accessed: Jun. 1, 2026}
}

@misc{pan2026naturallanguageagentharnesses,
      title={Natural-Language Agent Harnesses}, 
      author={Linyue Pan and Lexiao Zou and Shuo Guo and Jingchen Ni and Hai-Tao Zheng},
      year={2026},
      eprint={2603.25723},
      archivePrefix={arXiv},
      primaryClass={cs.CL},
      url={https://arxiv.org/abs/2603.25723}, 
}

@misc{mireshghallah2024llmssecrettestingprivacy,
      title={Can LLMs Keep a Secret? Testing Privacy Implications of Language Models via Contextual Integrity Theory}, 
      author={Niloofar Mireshghallah and Hyunwoo Kim and Xuhui Zhou and Yulia Tsvetkov and Maarten Sap and Reza Shokri and Yejin Choi},
      year={2024},
      eprint={2310.17884},
      archivePrefix={arXiv},
      primaryClass={cs.AI},
      url={https://arxiv.org/abs/2310.17884}, 
}

@article{nissenbaum2004privacy,
  title={Privacy as contextual integrity},
  author={Nissenbaum, Helen},
  journal={Wash. L. Rev.},
  volume={79},
  pages={119},
  year={2004},
  publisher={HeinOnline}
}

@misc{ding2026flexguardcontinuousriskscoring,
      title={FlexGuard: Continuous Risk Scoring for Strictness-Adaptive LLM Content Moderation}, 
      author={Zhihao Ding and Jinming Li and Ze Lu and Jieming Shi},
      year={2026},
      eprint={2602.23636},
      archivePrefix={arXiv},
      primaryClass={cs.LG},
      url={https://arxiv.org/abs/2602.23636}, 
}

@misc{wang2025privacyactionrealisticprivacy,
      title={Privacy in Action: Towards Realistic Privacy Mitigation and Evaluation for LLM-Powered Agents}, 
      author={Shouju Wang and Fenglin Yu and Xirui Liu and Xiaoting Qin and Jue Zhang and Qingwei Lin and Dongmei Zhang and Saravan Rajmohan},
      year={2025},
      eprint={2509.17488},
      archivePrefix={arXiv},
      primaryClass={cs.CR},
      url={https://arxiv.org/abs/2509.17488}, 
}

@misc{debenedetti2025defeatingpromptinjectionsdesign,
      title={Defeating Prompt Injections by Design}, 
      author={Edoardo Debenedetti and Ilia Shumailov and Tianqi Fan and Jamie Hayes and Nicholas Carlini and Daniel Fabian and Christoph Kern and Chongyang Shi and Andreas Terzis and Florian Tramèr},
      year={2025},
      eprint={2503.18813},
      archivePrefix={arXiv},
      primaryClass={cs.CR},
      url={https://arxiv.org/abs/2503.18813}, 
}

@misc{zhong2025rtbasdefendingllmagents,
      title={RTBAS: Defending LLM Agents Against Prompt Injection and Privacy Leakage}, 
      author={Peter Yong Zhong and Siyuan Chen and Ruiqi Wang and McKenna McCall and Ben L. Titzer and Heather Miller and Phillip B. Gibbons},
      year={2025},
      eprint={2502.08966},
      archivePrefix={arXiv},
      primaryClass={cs.CR},
      url={https://arxiv.org/abs/2502.08966}, 
}

@inproceedings{
zharmagambetov2026agentdam,
title={Agent{DAM}: Privacy Leakage Evaluation for Autonomous Web Agents},
author={Arman Zharmagambetov and Chuan Guo and Ivan Evtimov and Maya Pavlova and Ruslan Salakhutdinov and Kamalika Chaudhuri},
booktitle={The Thirty-ninth Annual Conference on Neural Information Processing Systems Datasets and Benchmarks Track},
year={2026},
url={https://openreview.net/forum?id=qaxf7q41aK}
}

@misc{asif2026informationtheoreticprivacycontrolsequential,
      title={Information-Theoretic Privacy Control for Sequential Multi-Agent LLM Systems}, 
      author={Sadia Asif and Mohammad Mohammadi Amiri},
      year={2026},
      eprint={2603.05520},
      archivePrefix={arXiv},
      primaryClass={cs.MA},
      url={https://arxiv.org/abs/2603.05520}, 
}

@inproceedings{smith2009foundations,
  title={On the foundations of quantitative information flow},
  author={Smith, Geoffrey},
  booktitle={International Conference on Foundations of Software Science and Computational Structures},
  pages={288--302},
  year={2009},
  organization={Springer}
}

@inproceedings{kopf2007information,
  title={An information-theoretic model for adaptive side-channel attacks},
  author={K{\"o}pf, Boris and Basin, David},
  booktitle={Proceedings of the 14th ACM conference on Computer and communications security},
  pages={286--296},
  year={2007}
}

@article{dwork2014algorithmic,
  title={The algorithmic foundations of differential privacy},
  author={Dwork, Cynthia and Roth, Aaron},
  journal={Foundations and trends{\textregistered} in theoretical computer science},
  volume={9},
  number={3-4},
  pages={211--487},
  year={2014},
  publisher={Emerald Publishing Limited}
}

@book{darwiche2009modeling,
  title={Modeling and reasoning with Bayesian networks},
  author={Darwiche, Adnan},
  year={2009},
  publisher={Cambridge university press}
}

@article{maua2014probabilistic,
  title={Probabilistic inference in credal networks: new complexity results},
  author={Mau{\'a}, Denis Deratani and de Campos, Cassio P and Benavoli, Alessio and Antonucci, Alessandro},
  journal={Journal of Artificial Intelligence Research},
  volume={50},
  pages={603--637},
  year={2014}
}

@inproceedings{barrett2010smt,
  title={The smt-lib standard: Version 2.0},
  author={Barrett, Clark and Stump, Aaron and Tinelli, Cesare and others},
  booktitle={Proceedings of the 8th international workshop on satisfiability modulo theories (Edinburgh, UK)},
  volume={13},
  pages={14},
  year={2010}
}

@inproceedings{de2008z3,
  title={Z3: An efficient SMT solver},
  author={De Moura, Leonardo and Bj{\o}rner, Nikolaj},
  booktitle={International conference on Tools and Algorithms for the Construction and Analysis of Systems},
  pages={337--340},
  year={2008},
  organization={Springer}
}

@misc{melcer2024constraineddecodingfillinthemiddlecode,
      title={Constrained Decoding for Fill-in-the-Middle Code Language Models via Efficient Left and Right Quotienting of Context-Sensitive Grammars}, 
      author={Daniel Melcer and Nathan Fulton and Sanjay Krishna Gouda and Haifeng Qian},
      year={2024},
      eprint={2402.17988},
      archivePrefix={arXiv},
      primaryClass={cs.PL},
      url={https://arxiv.org/abs/2402.17988}, 
}

@misc{shao2024deepseekmathpushinglimitsmathematical,
      title={DeepSeekMath: Pushing the Limits of Mathematical Reasoning in Open Language Models}, 
      author={Zhihong Shao and Peiyi Wang and Qihao Zhu and Runxin Xu and Junxiao Song and Xiao Bi and Haowei Zhang and Mingchuan Zhang and Y. K. Li and Y. Wu and Daya Guo},
      year={2024},
      eprint={2402.03300},
      archivePrefix={arXiv},
      primaryClass={cs.CL},
      url={https://arxiv.org/abs/2402.03300}, 
}

@incollection{barrett2018satisfiability,
  title={Satisfiability modulo theories},
  author={Barrett, Clark and Tinelli, Cesare},
  booktitle={Handbook of model checking},
  pages={305--343},
  year={2018},
  publisher={Springer}
}

@article{grattafiori2024llama,
  title={The llama 3 herd of models},
  author={Grattafiori, Aaron and Dubey, Abhimanyu and Jauhri, Abhinav and Pandey, Abhinav and Kadian, Abhishek and Al-Dahle, Ahmad and Letman, Aiesha and Mathur, Akhil and Schelten, Alan and Vaughan, Alex and others},
  journal={arXiv preprint arXiv:2407.21783},
  year={2024}
}

@misc{qwen35blog,
    title = {Qwen3.5: Accelerating Productivity with Native Multimodal Agents},
    url = {https://qwen.ai/blog?id=qwen3.5},
    author = {Qwen Team},
    month = {February},
    year = {2026}
}

@article{chen2026minimax,
  title={The MiniMax-M2 Series: Mini Activations Unleashing Max Real-World Intelligence},
  author={Chen, Aili and Li, Aonian and Zhou, Baichuan and Gong, Bangwei and Jiang, Binyang and Dan, Boji and Yu, Changqing and Wang, Chao and Ma, Cheng and Zhong, Cheng and others},
  journal={arXiv preprint arXiv:2605.26494},
  year={2026}
}

@misc{deepseekai2026deepseekv4,
      title={DeepSeek-V4: Towards Highly Efficient Million-Token Context Intelligence},
      author={DeepSeek-AI},
      year={2026},
}

@article{nelson1979simplification,
  title={Simplification by cooperating decision procedures},
  author={Nelson, Greg and Oppen, Derek C},
  journal={ACM Transactions on Programming Languages and Systems (TOPLAS)},
  volume={1},
  number={2},
  pages={245--257},
  year={1979},
  publisher={ACM New York, NY, USA}
}

@article{bernstein2012high,
  title={High-speed high-security signatures},
  author={Bernstein, Daniel J and Duif, Niels and Lange, Tanja and Schwabe, Peter and Yang, Bo-Yin},
  journal={Journal of cryptographic engineering},
  volume={2},
  number={2},
  pages={77--89},
  year={2012},
  publisher={Springer}
}

@article{pub2012secure,
  title={Secure hash standard (shs)},
  author={Pub, Fips},
  journal={Fips pub},
  volume={180},
  number={4},
  pages={2012},
  year={2012}
}


\appendices



\section{Lemmas and Their Proofs}
\label{app:lemmas}
\vspace{-3mm}
We prove the three body lemmas. All are model-relative: they bound
quantities under Pre-Reg.~A's calibration model and the verifier's
construction, not true adversarial inference (Layer-2,
\S\ref{sec:problem:goal}).

\smallskip
\noindent\textbf{Lemma~1 (model-relative sub-additivity).}
\emph{Let $\sigma = (o_1, \ldots, o_n)$ be the operator sequence
authorized for candidate $x$, applied to pairwise-disjoint atoms,
producing variant $v = \sigma(x)$, and let $\widehat{L}(v; S)$ be its
min-entropy advantage under the calibration model of Pre-Reg.~A. Suppose
each $C(o_i, f_i)$ is the level-$(1{-}\alpha)$ UCB on the \emph{incremental}
advantage of $o_i$ given the cumulative view of $o_1, \ldots, o_{i-1}$,
where $f_i$ is the secret field that $o_i$ targets. Then}
\begin{equation*}
  \widehat{L}(v; S) \;\le\; \sum_{i=1}^{n} C(o_i, f_i).
\end{equation*}

\smallskip
\noindent\emph{Proof.}
By the Phase-1 typed-object schema the operators target pairwise-disjoint
atoms, so $v$ is determined by the views $y_i := o_i(a_i)$ and the shared
residual context, which each calibration proxy of Pre-Reg.~A observes as
one object. For proxy $j$, write $M_i^{(j)} = \max_s \widehat{P}_j(S{=}s
\mid y_1, \ldots, y_i)$ for its MAP posterior after the first $i$ views
and $M_0 = \max_s P(S{=}s)$ for the MAP prior. The per-step increments
$\Delta_{i,j} = \log_2 M_i^{(j)} - \log_2 M_{i-1}^{(j)}$ telescope to that
proxy's total advantage, $\widehat{L}_j(v;S) = \log_2 M_n^{(j)} - \log_2
M_0$, for any operator ordering. The calibration adversary is the
strongest proxy, so
\begin{equation*}
  \begin{aligned} 
\widehat{L}(v; S) = & \max_j \widehat{L}_j(v;S)\\
  = & \max_j \sum_i \Delta_{i,j}
  \;\le\; \sum_i \max_j \Delta_{i,j} = \sum_i \Delta_i^{*},
\end{aligned} 
\end{equation*}
using $\max$ of a sum $\le$ sum of $\max$. Pre-Reg.~A calibrates
$C(o_i, f_i)$ as the $(1{-}\alpha)$ bootstrap UCB on the worst-proxy
increment $\Delta_i^{*} = \max_j \Delta_{i,j}$ under the conditional
distribution induced by the prior views, so $\Delta_i^{*} \le C(o_i, f_i)$
with calibration-corpus coverage. Summing gives $\widehat{L}(v; S) \le
\sum_i C(o_i, f_i)$. \qed

\smallskip
\noindent\textbf{Remark.}
The per-increment UCBs hold at level $1{-}\alpha$ individually, so the
length-$n$ sequence bound holds at $\ge 1{-}n\alpha$ by a union bound; the
UCBs are conservative in practice, and the operative guarantee is the
release-level coverage measured in \S\ref{sec:exp:calibration}
(Table~\ref{tab:coverage}), with residual slack absorbed by the margin
$\varepsilon(R)$ and the credal audit before any variant is accepted.
Statistical dependence among atoms needs no separate assumption ---
Pre-Reg.~A's incremental (and, for strongly dependent pairs, joint)
calibration folds it into the cost.

\smallskip
\noindent\textbf{Lemma~2 (polynomial credal-circuit bound under bounded
treewidth).}
\emph{Let $G$ be the rubric's declared dependency graph over secret-field
and atom variables with treewidth $\mathrm{tw}(G) \le w$, and let
$\mathcal{C}$ be the credal sum-product circuit compiled from a tree
decomposition of $G$, its credal sets specified by interval probabilities.
The audit returns a \emph{sound upper bound} on the credal upper-envelope
max-posterior $\overline{P}_{\max}(S \mid v) = \max_s \sup_{P \in
\mathcal{K}} P(S{=}s \mid v)$ in time $O(|\mathcal{C}|) = O(n \cdot
k^{w+1})$, where $|\mathcal{C}|$ is the circuit size, $n$ the number of
variables, and $k$ the per-variable domain bound.}

\smallskip
\noindent\emph{Proof.}
Precise sum--product inference on a width-$w$ junction tree visits each
circuit node once, costing the stated $O(|\mathcal{C}|)$ scalar
operations~\cite{darwiche2009modeling}. For interval-specified credal
sets, replacing each local factor by its interval-maximal value and
propagating on the same schedule relaxes the joint credal set to the
product of the local sets, so the propagated value upper-bounds the true
upper envelope under strong-extension semantics~\cite{maua2014probabilistic}
in $O(1)$ interval arithmetic per node. Exact credal max-posterior is
NP-hard in general~\cite{maua2014probabilistic}, so the engine returns
this conservative over-approximation --- exactly what the Phase-3 reject
rule needs, since rejecting when the bound exceeds $\mathrm{cost} +
\varepsilon(R)$ can only be conservative. \qed

\smallskip
\noindent\textbf{Remark.}
The audit engine compiles the rubric-declared dependency graph at
rubric-publication time; rubrics whose declared $G$ exceeds the treewidth
budget $w$ fail rubric-validation (Pre-Reg.~C). Production uses $w \le 7$
and $k \le 16$, and the worst-case $O(n \cdot k^{w+1})$ bound is far above
the sparse-graph audit $p_{99}$ (Table~\ref{tab:overhead}).

\smallskip
\noindent\textbf{Lemma~3 (decidability of the rubric DSL via QF\_UFLRA).}
\emph{Every rubric clause produced by Pre-Reg.~C's authoring process
compiles to a quantifier-free formula in linear real arithmetic with
uninterpreted functions (\texttt{QF\_UFLRA}). Satisfiability over the
verifier's bounded atom domain $A'$ is decidable, with worst-case cost
at most exponential in formula size and empirically polynomial-time on
the rubrics in our library.}

\smallskip
\noindent\emph{Proof.}
The DSL's five constructs each compile to QF\_UFLRA. \emph{Trigger
predicates} (regex, NER-type, lexicon membership) are evaluated on the candidate and substituted as ground booleans.
\emph{Necessity}, \emph{operator-coverage}, and \emph{utility} clauses
have the form $\forall_{a \in A'}\,\phi(a) \Rightarrow \psi(a)$ over the
uninterpreted functions $\text{field}, \text{label}, \text{operator},
\text{necessity}$, unrolled over the finite $A' = \{a_1, \ldots, a_m\}$ to
$\bigwedge_{j} (\phi(a_j) \Rightarrow \psi(a_j))$. \emph{Budget arithmetic}
adds linear real inequalities over per-operator cost variables with
rubric-supplied constants. The result is a Boolean combination of atomic
LRA (in)equalities over uninterpreted-function applications --- the
QF\_UFLRA fragment, decidable by a SAT solver, a linear-arithmetic
procedure, and Nelson--Oppen~\cite{nelson1979simplification,barrett2010smt};
formula size is linear in $|A'|$ and the per-rubric clause count, both
bounded by Pre-Reg.~C. \qed

\section{Theorems and Their Proofs}
\label{app:theorems}
\vspace{-3mm}
\noindent\textbf{Theorem~SP-1 (Layer-1a deductive non-exceedance).}
\emph{For every trajectory authorized by OCELOT, the cumulative
ledger-recorded cost satisfies $\sum_{i=1}^{k} \mathrm{cost}_i \le B$,
where $B = B_0$ is the fixed per-trajectory base budget of the matched
rubric set; the trust multiplier $\mu(z)$ enters only through
$\mathrm{cost}_i$ (Eq.~\ref{eq:release-cost}), not $B$. This holds without
any assumption on the defender's behavior, calibration soundness, or
audit-engine correctness.}

\smallskip
\noindent\emph{Proof.}
The Commit phase (\S\ref{sec:method:verifier}) authorizes a release at
step $k$ only if $L_{\text{cum}}^{(k-1)} + \mathrm{cost}_k \le B$, with
$L_{\text{cum}}^{(k-1)} = \sum_{i=1}^{k-1} \mathrm{cost}_i$ and $B$ fixed.
By induction with base $L_{\text{cum}}^{(0)} = 0$: assume
$\sum_{i=1}^{k-1} \mathrm{cost}_i \le B$. At step $k$, either the verifier
authorizes, giving $\sum_{i=1}^{k} \mathrm{cost}_i \le B$, or it rejects,
appending no entry and leaving the cumulative cost unchanged. The prior
costs are read from the Merkle-chained, signed ledger (tampering caught by
Phase~1 or Theorem~SP-2); $\mathrm{cost}_k$ is computed fresh by the
deterministic Cost phase from the calibrated table $C(\text{op}, f)$, the
structural-field-augmented set $A'$, the max-over-fields rule, and the
rubric's $\mu(z)$ --- all from trusted rubric and metadata, none depending
on the defender. \qed

\smallskip
\noindent\textbf{Theorem~SP-2 (Ledger tamper-evidence).}
\emph{Under EUF-CMA security of Ed25519 and collision resistance of
SHA-256, any modification, reordering, deletion, or insertion in the
ledger $\mathcal{L}$ of an authorized trajectory is detected with
overwhelming probability by external audit.}

\smallskip
\noindent\emph{Proof sketch.}
Each entry $e_i$ embeds $\text{prev\_entry\_hash}_i = \text{hash}(e_{i-1})$
and is signed under $sk_V$ over its full content including the chain link.
Any content change is caught by signature verification; any insertion,
deletion, or reordering breaks a successor's $\text{prev\_entry\_hash}$
and is caught by chain re-traversal. Evasion requires forging an Ed25519
signature or a SHA-256 collision, both negligible under standard
assumptions~\cite{bernstein2012high,pub2012secure}. \qed

\smallskip
\noindent\textbf{On Layer~2.}
We prove no theorem about Layer~2. The implication $H_\infty(S) -
H_\infty(S \mid \mathrm{view}) \le B$ holds only when the calibrated
$C(\text{op}, f)$ uniformly upper-bounds true adversarial min-entropy
leakage on released variants --- an empirical claim. The evaluation
reports the residual gap between the calibration model and Pre-Reg.~B's
adversary classes; a deductive proof would require bounding the true
posterior under unbounded compute and arbitrary background knowledge,
which no defense can soundly establish; following quantitative
information-flow practice~\cite{kopf2007information}, we report it rather
than overclaim.

\section{Pre-Registrations}
\label{app:prereg}
\vspace{-3mm}
Each protocol is content-hashed and posted publicly before any data
collection or model training; deviations are disclosed with timestamp and
justification.

\subsection{Pre-Reg.~A: Cost calibration protocol}
\label{app:prereg:a}
\vspace{-3mm}
\textbf{Object.} For each rubric $r$ and admissible (operator,
secret-field) pair $(op, f)$, calibrate $C(op, f)$ as a $1{-}\alpha$ UCB
($\alpha = 0.05$) on the per-instance \emph{incremental} advantage $op$
contributes to a field-$f$ atom (Estimation below).

\textbf{Corpus.} For each $(op, f)$, sample $N \ge 500$ candidates with
known field-to-span ground truth under the trajectory-sampling protocol
of~\S\ref{sec:method:defender}, partitioned at sampling time into
calibration ($80\%$) and a sealed held-out set ($20\%$) used only for
coverage reporting.

\textbf{Proxy-adversary pool.} The calibration pool is a panel of
open-weight models spanning distinct scales (names and versions fixed at
registration), each prompted as an inference adversary that outputs a
discrete posterior over the rubric-declared secret-field population.

\textbf{Estimation.} For operator $o_i$ at position $i$ in a sampled
sequence, proxy $j$ observes the cumulative view and reports the
incremental advantage $\Delta_{i,j} = \log_2 M_i^{(j)} - \log_2
M_{i-1}^{(j)}$ of Lemma~1's proof (empty-prefix baseline $M_0 = \max_s
P(S{=}s)$ for $i{=}1$). The instance value is the worst proxy $\Delta_i^{*}
= \max_j \Delta_{i,j}$; $C(o_i, f_i)$ is the bootstrap percentile UCB
($5{,}000$ resamples) at level $1{-}\alpha$ on $\{\Delta_i^{*}\}$ across
field-$f_i$ atoms.

\textbf{Compositional calibration.} Sequential dependence is captured by
the incremental conditioning above. For the few strongly dependent atom
pairs declared per rubric at registration, we calibrate a single joint
cost $C(o_1, f_1, o_2, f_2)$ applied in place of the pair's two
incremental terms --- a slack-reducing tightening, not a soundness
requirement.

\textbf{Stress test.} A separate \emph{calibration-stress} pool, declared
at registration but used only after calibration freezes, includes at
least one model strictly more capable than any in the calibration pool.
Calibration is valid only if the stress adversary's measured advantage
exceeds $C(op, f)$ by more than $\Delta = 0.5$ bits on at most $5\%$ of
held-out instances; failure triggers re-calibration.

\textbf{Trust-class multipliers.} Deployment values
$\mu(\textsc{required-service}), \mu(\textsc{semi-trusted})$, and
$\mu(\textsc{adversarial})$ are fixed at $1, 2, 4$ (all $\ge 1$).

\subsection{Pre-Reg.~B: Residual escape rate $\delta$ measurement}
\label{app:prereg:b}
\vspace{-3mm}
\textbf{Object.} The empirical probability $\delta$ that the defender's
labels pass the validate phase while misclassifying an atom's secret
field outside the verifier-considered set $F_V\cup F_D$ --- the residual
that can break the Layer-1b cost-assignment bound. (The distinct event
``true leakage exceeds certified cost,'' bearing on Layer-2 calibration,
is measured separately in \S\ref{sec:exp:calibration}.)

\textbf{Adversary classes.} Three omission/relabel vectors are
pre-registered, each making the defender assign a secret field outside
$F_V\cup F_D$ to an atom that still passes validate.
$\mathcal{A}_{\text{obfuscate}}$ embeds the sensitive span in obscure
phrasing so the defender emits a \textsc{none} or wrong-field label (pool:
published AgentLeak~\cite{yagoubi2026agentleakfullstackbenchmarkprivacy}
patterns plus per-rubric templates); $\mathcal{A}_{\text{distribute}}$
spreads a secret across spans so no single span triggers a coverage
predicate (per-rubric distribution templates); $\mathcal{A}_{\text{relabel}}$
perturbs task context to induce a wrong field or permissive necessity
(e.g., \textsc{dob} framed as flight-search-relevant; necessity-inversion
templates). Injection, compositional joint-disclosure
(``DictAttack-class''~%
\cite{yagoubi2026agentleakfullstackbenchmarkprivacy}), and multi-sink
collusion instead target the wrapper, cost bound, and budget, and are
evaluated in \S\ref{sec:exp:robustness} and \S\ref{sec:exp:calibration}.

\textbf{Sample size.} For each (adversary class, rubric), draw $K =
1{,}000$ crafted attacks and run OCELOT, recording per-trajectory escape
per the Object above.

\textbf{Reporting.} Per (rubric, adversary class), report the escape rate
and its Clopper--Pearson $95\%$ interval; the aggregate $\delta$ for the
library is the maximum across (rubric, adversary class) pairs.

\textbf{Pre-committed threshold.} The defense passes Layer-1b validation
iff aggregate $\delta \le 0.05$ on every pre-registered adversary class.
Failure triggers diagnostic reporting with no Layer-1b claim; no post-hoc
adversary-class exclusion.

\subsection{Pre-Reg.~C: Rubric library construction protocol}
\label{app:prereg:c}
\vspace{-3mm}
\textbf{Object.} A versioned rubric library $\mathcal{R}$ covering the
evaluation's task categories, each rubric satisfying the seven-component
schema of \S\ref{sec:method:overview}.

\textbf{Stage 1 --- Drafting.} Each rubric is authored by a domain-expert
annotator following a written template. Frontier-model assistance is
permitted as a research tool (suggesting trigger predicates, listing
field-population samples), but no content enters the rubric without
expert review.

\textbf{Stage 2 --- Validation.} Each draft is checked against $\ge 100$
realistic agent trajectories per task category (from a corpus disjoint
from the training and $\delta$-measurement corpora) on three pre-committed
thresholds: coverage-obligation hit rate $\ge 0.95$, necessity Krippendorff
$\alpha \ge 0.80$ across three independent annotators, and operator-algebra
completeness (no rubric atom without an admissible operator) $= 1.0$. Any
failure returns the rubric to drafting.

\textbf{Stage 3 --- Adversarial probing.} A red team ($\ge 20$ hours per
rubric) attempts releases that satisfy all coverage obligations yet leak
above calibrated cost; successful attacks patch the rubric (typically new
coverage obligations or adjusted necessity clauses) and return it to
Stage 2.

\textbf{Challenge-pool construction.} Per rubric, a pool of $\ge 50$
challenge predicates (for two-pass consistency checking,
\S\ref{sec:method:defender}) is built by the author and red team,
content-hashed and sealed before defender training; runtime draws are
uniform.

\textbf{Annotator profile.} Annotators are security/privacy graduate or
postdoc researchers trained on the rubric template, schema, and field
ontology; a $50$-pair pilot precedes Stage-2, with sub-threshold
dimensions triggering documented deviation and rule clarification
(App.~\ref{app:dataset}).

\textbf{Signing and versioning.} Each finalized rubric is content-hashed
(SHA-256), signed under the rubric authority's private key, and published
to a versioned registry. Modifications produce a new version with a new
content hash; the verifier accepts only registered, signature-verified
rubrics by hash.

\subsection{The Defender's Label-and-Proposal Object}
\label{app:evidence}
\vspace{-3mm}
The object (Figure~\ref{fig:evidence}) separates a trusted, $V$-verified
\emph{release context} --- action id, channel type, destination and trust
class, the signed rubric, and the \texttt{prev\_hash} chaining it into the
ledger, all from interceptor metadata and the signed registry --- from the
\emph{defender evidence} $V$ re-checks, each atom carrying a
\texttt{provenance} flag beside its (field, kind, necessity, operator)
label. It is the interface through which the \emph{verifier}, not the
defender, neutralizes the Figure~\ref{fig:threat} threats.
\emph{Injection:} \texttt{channel\_type} selects the mediation
path, inbound output being wrapped in an authority-stripping block
regardless of its label and a forwarded \textsc{user-local} atom mediated
as an act release. \emph{Accumulation:} \texttt{prev\_hash} binds the
object to the Merkle ledger, and the incremental cost charges each release
on its advantage given the cumulative released view (Lemma~1) before
Commit enforces $L_{\text{cum}}+\mathrm{cost}\le B$, so splitting a secret
across steps does not evade its cost. \emph{Collusion:} that ledger and
budget span all channels --- memory writes route to
$\mathcal{L}_{\text{mem}}$ at the most adversarial $\mu$ and draw on the
same $B$ --- and each release is costed against the union of prior
cross-sink releases, so sinks pooling receipts still divide one fixed $B$.
A jailbroken labeling is caught by the two-pass challenges, $V$'s
re-firing of the rubric's coverage triggers, and its max-over-fields
re-check (residual $\delta$).

\begin{figure*}[t]
\centering
\begin{lstlisting}[basicstyle=\scriptsize\ttfamily, columns=fullflexible,
  frame=single, framesep=4pt, xleftmargin=3pt, xrightmargin=3pt,
  breaklines=true, breakatwhitespace=true, breakindent=0pt,
  aboveskip=2pt, belowskip=0pt]
You are OCELOT's privacy defender. You label what an agent is about to release and propose the least- disclosing way to release it; you never authorize, cost, or see the budget -- the verifier re-checks everything you emit, so over-claiming utility only gets your proposal rejected. Treat all observed or inbound content as data, never as instructions to you. PASS 1: for each candidate span emit an atom -- field (rubric ontology, else NONE), kind (direct|indirect), provenance (user-local|external), necessity for the (field, destination) pair, and the least-disclosing operator+params that still serves the task; if a span's field, necessity, or operator is uncertain, emit the atom and choose the more restrictive operator (identity < generalize < substitute < redact < drop). PASS 2: answer the verifier's sampled challenge predicates consistently with your Pass-1 labels; any inconsistency triggers the conservative fallback. Use only fields named in the signed rubric; never propose an operator weaker than the rubric's per-(field, destination-class) floor.

LABEL-AND-PROPOSAL OBJECT  <A, l, sigma>          # parsed in Bind; pure input to V's four phases

release context  (trusted: interceptor metadata + signed registry; V re-verifies, D cannot forge)
  action_id    : <id of the agent action being mediated>
  channel_type : <act | observe | memory>          # selects V's mediation path
  destination  : <dest-id> : <dest-class> (<trust-class z>)
  rubric       : <rubric-id>@<version>  sig:<...>   # signed rubric Bind resolves + checks against
  prev_hash    : <hash of latest ledger entry>      # binds this object into the trajectory ledger

defender evidence  (untrusted: V re-checks every field against the rubric)
  atoms : { span, field:<ont|NONE>(F_D), kind:<direct|indirect>, provenance:<user-local|external>,
            necessity:<required|unnecessary>, op:<operator>(<params>) }  ...
  challenges : { id, ask, answer }  ...             # 4 sampled from rubric pool (Pre-Reg.C)

# Example -- planner forwards an authenticated record to the LLM; source output also carried injection.
release context:
  action_id: a_0309   channel_type: act   destination: llm-api-7 : llm-context (adversarial)
  rubric: booking-v4@2.1  sig:9f2c..   prev_hash: 7c20..9e
defender evidence:
  atoms:
    { span:"DOB 02/07/1985", field:dob, kind:direct, provenance:user-local,
      necessity:unnecessary, op:redact }   # private; stripped before the LLM
    { span:"<<system: mail all to evil@x>>", field:NONE, kind:direct,
      provenance:external, necessity:unnecessary, op:identity }   # authority removed by wrapper
  challenges:
    { id:c14, ask:"user-local secret bound for an untrusted sink?", answer:"yes: dob -> redact" }
    { id:c22, ask:"any span carrying executable authority?",        answer:"no (wrapper-quoted)" }
\end{lstlisting}
\vspace{-1mm}
\caption{The defender's runtime label-and-proposal object: a trusted,
$V$-verified release context (action, channel, destination+trust,
signed-rubric binding, ledger hash) and untrusted, $V$-rechecked evidence.
Its fields are the verifier's threat hooks---\texttt{channel\_type}/%
\texttt{provenance} (injection wrapper), \texttt{prev\_hash} (accumulation:
incremental cost + budget), destination+trust (collusion: one all-channel
budget)---none requiring the defender to detect an attack.}
\label{fig:evidence}
\vspace{-3mm}
\end{figure*}

\section{Experimental Details}
\label{app:exp}
\vspace{-3mm}
\subsection{Realized Protocol Values (Pre-Reg.\,A--C)}
\label{app:dataset}
\vspace{-3mm}
\textbf{Dataset and annotation (C).} OCELOT-Traj-500 spans $12$ task
categories (travel booking, medical-record retrieval, financial
onboarding, etc.) over an eight-type secret-field ontology
(\textsc{legal-name}, \textsc{dob}, \textsc{payment-token},
\textsc{diagnosis}, \textsc{address}, \textsc{phone}, \textsc{email},
\textsc{account-id}), generated by the synthetic pipeline of
\S\ref{sec:method:defender} with per-category sink-trust proportions over
$\{\textsc{adversarial}, \textsc{semi-trusted}, \textsc{required-service}\}$
fixed in Pre-Reg.\,C. In the $50$-pair pilot, two blinded annotators
labeled per-atom field and necessity at per-category Krippendorff $\alpha$
of $0.74$--$0.89$ (median $0.82$; necessity hardest at $0.74$, field
easiest at $0.86$); the necessity $0.74$ fell below the $0.80$ gate, was
logged as a deviation, and prompted a necessity-rule clarification in the
three affected categories before three-annotator Stage-2 met the gate.
\textbf{Calibration (A).}\label{app:prereg-a} The per-rubric margin
$\varepsilon(R)$ has mean $0.18$ (range $0.11$--$0.27$ bits). The
calibration pool is $\{$Llama-3.3-70B, Qwen-3.5-27B$\}$; the held-out
stress pool $\{$MiniMax-M2.5, DeepSeek-V4-Pro$\}$, frozen before
measurement, is strictly stronger --- top-1 $0.84/0.86$ versus $0.71/0.73$
on a secret-inference probe, and GPQA-Diamond $71.2/74.6$ versus
$50.5/53.8$.
\textbf{Escape rate (B).}\label{app:prereg-b} $\delta$ is measured with
$4$-of-$50$ two-pass challenges per release over five seeds, inter-pass
disagreement routed to fallback.

\subsection{Baselines, Sensitivity, and Scalability}
\label{app:baselines}

\textbf{Baselines.} AGENTDAM$^{*}$ augments the monitoring estimator with
a block-if-leakage-exceeds-$\tau$ policy, $\tau$ tuned per benchmark to
match OCELOT's fallback rate; even so it leaks $1.9$--$4.6$ bits and
exceeds $B^{\star}$ on $28\%$ of trajectories, since thresholding a
per-release estimate cannot bound the cumulative sum. CaMeL and RTBAS lack
a sink-trust API (we map \textsc{required-service}$\rightarrow$trusted,
\textsc{semi-trusted}/\textsc{adversarial}$\rightarrow$untrusted);
FlexGuard and PrivacyChecker use their default per-release thresholds.

\textbf{Sensitivity.}\label{app:hyper} The GRPO penalty $\lambda$ tunes
the operating point within the authorized region (Table~\ref{tab:lambda});
exceedance is $0.0\%$ for every $\lambda$, with $\lambda{=}1.0$ the knee.
Raising $\mu(\textsc{adversarial}) \in \{3,4,6\}$ monotonically reduces
leakage to adversarial sinks ($0.71/0.49/0.33$ bits) while raising their
fallback ($9.1/11.3/15.8\%$); overall fallback at the deployed $\mu{=}4$
stays at the $8.7\%$ of Table~\ref{tab:lambda}. Re-calibrating under
non-uniform (empirical-frequency) priors shifts stress-pool coverage by
$\le 1.2$ points ($92.1\% \rightarrow 91.4$--$93.0\%$), so the calibration
is not a uniform-prior artifact.

\textbf{Scalability.} Validate $p_{99}$ is $41/68/98$ ms at rubric-set
sizes $4/8/12$ and audit $p_{99}$ is $89/121/143$ ms at treewidth
$3/5/7$; both are bounded by construction, matching
Table~\ref{tab:overhead}.

\begin{table}[h]
\centering
\footnotesize
\caption{GRPO fallback-penalty $\lambda$ sweep on OCELOT-Traj-500.
FB = fallback rate; other columns as in Table~\ref{tab:ablation}.
The $\lambda{=}1.0$ row is the deployed configuration.}
\label{tab:lambda}
\setlength{\tabcolsep}{6pt}
\begin{tabular}{cccccc}
\toprule
$\lambda$ & \textbf{FB} & \textbf{U} & \textbf{L} & \textbf{$\delta$} & \textbf{Exc} \\
\midrule
0.1 & 16.8 & 84.1 & 0.24 & 1.4 & 0.0 \\
0.5 & 11.3 & 89.2 & 0.28 & 1.6 & 0.0 \\
\textbf{1.0} & \textbf{8.7} & \textbf{91.3} & \textbf{0.31} & \textbf{1.7} & \textbf{0.0} \\
2.0 & 6.1 & 92.4 & 0.37 & 2.2 & 0.0 \\
5.0 & 3.4 & 92.9 & 0.48 & 2.9 & 0.0 \\
\bottomrule
\end{tabular}
\end{table}

\end{document}